\documentclass[12pt]{article}
\usepackage[dvips]{graphicx}  

\begin{document}

\begin{center}
{\Large \bf Unravelling the  Dark Matter  - Dark Energy Paradigm  } 

\bigskip

Reginald T. Cahill 

{\small \it  School of Chemistry, Physics and Earth Sciences, Flinders University,
Adelaide 5001, Australia}\\

{\footnotesize  Reg.Cahill@flinders.edu.au}

\begin{abstract}
The standard  $\Lambda$CDM  model of cosmology  is usually understood to arise from demanding that the Friedmann-Lema\^{i}tre-Robertson-Walker  (FLRW) metric satisfy   the  General Relativity  dynamics for spacetime metrics.  The FLRW    data-based dominant parameter values, $\Omega_\Lambda=0.73$ and  $\Omega_m=0.27$  for the dark energy and dark matter+matter,  respectively, are then determined by fitting the supernova red-shift data. However  in the  pressure-less flat-space case  the $\Lambda$CDM  model is most easily derived from Newtonian gravity, and which was based on the special case of planetary motion in the solar system. Not surprisingly when extended to galactic rotations and cosmology  Newtonian dynamics  is found to be wanting, and the fix-up involves introducing dark matter and dark energy, as shown herein.  However a different theory of gravity leads to a different account of galactic rotations and  cosmology, and does not require dark matter nor dark energy to fit the supernova data.  It is shown that fitting the $\Lambda$CDM  model to this new model, and so independently  of the actual supernova data, requires the $\Lambda$CDM  model parameters to be those given above. Hence we conclude that dark energy and dark matter are no more than mathematical artifacts to fix-up  limitations of Newtonian gravity.   Various other data are also briefly reviewed to    illustrate other successful tests of this new theory of gravity.

\end{abstract}
\end{center}

PACS: {98.80.-k, 98.80.Es, 95.35.+d, 95.36.+x


\newpage


\section{Introduction}

The current  $\Lambda$CDM  standard model of cosmology is based upon General Relativity (GR) as applied to the spatially-flat Friedmann-Lema\^{i}tre-Robertson-Walker  (FLRW-GR) spacetime metric  together with  the Weyl postulate for the energy-momentum density tensor, leading to the Friedmann equations for the 3-space scale factor \cite{Friedmann, Lemaitre, Robertson, Walker}\footnote{We use FLRW-GR as a full acronym for the model, as in section 14 the FLRW metric arises in a non-GR context. }.  Fitting this model to the magnitude-redshift data from supernovae and gamma-ray-burst (GRB) data requires the introduction of dark energy and  dark matter, and a concomitant future exponential acceleration of the universe \cite{PS}.  The dark energy has been most simply interpreted as a cosmological constant $\Lambda$.  Fitting the data gives $\Omega_\Lambda=0.73$ and  $\Omega_{m}=0.27$, with baryonic matter 
forming only some $\Omega_b=0.05$ of  $\Omega_{m}$, so that the `dark matter' component has $\Omega_{DM}=0.22$.  Hence according to the FLRW-GR model the universe expansion is determined   mainly by dark energy and cold dark matter, leading to the $\Lambda$CDM label.  A peculiar aspect of the  $\Lambda$CDM model is that the universe can {\it only} expand if the energy density is non-zero, i.e. space itself  cannot  expand without that energy density being present.  This has been a feature of the FLRW-GR dynamics from the beginning of cosmology, and as shown herein is a direct consequence of extending Newtonian gravity to cosmology, and so well beyond its established regime.  It is probably not well known that the $\Lambda$CDM model is a simple and direct consequence of Newtonian gravity, as shown later. Here we derive a new cosmology which leads to, apart from other numerous tests, an expanding flat 3-space which does not require the presence of energy for that expansion.  This expansion gives a parameter-free fit to the supernovae/GRB data, without invoking dark energy or dark matter.  Nevertheless, if we best-fit the  FLRW-GR  $\Lambda$CDM model to the new cosmology dynamics over the redshift range $z\in\{0,14\}$,  by varying $\Omega_\Lambda$, we obtain   $\Omega_\Lambda=0.73$, $\Omega_{m}=1-\Omega_\Lambda=0.27$. In other words, if the new cosmological model is valid, then we can predict that fitting the $\Lambda$CDM model to the data will give the parameter values exactly as reported.
However the new cosmology does not predict an accelerating universe; that is merely a spurious consequence of the FLRW-GR model having the wrong functional form for  its Hubble function.   These results change completely our understanding of the evolution of the universe, and of its contents. Basically there is just a very small amount of conventional matter, as indeed deduced from CMB temperature fluctuation data, and a dominant   expanding dynamical 3-space.

\section{The $\Lambda$CDM Model from  Newtonian Gravity}
The simplest and most direct derivation of a theoretical model is also usually the most instructive and most revealing, for abstract formalism is very effective at hiding  fundamental issues. Here we derive the $\Lambda$CDM model directly and simply from Newtonian gravity\footnote{This derivation has a long history, but appears to have sunk without trace in the  context of the dark energy and dark matter debate. }.  Newtonian gravity was based on Kepler's observations of the motion of planets within the solar system, with the attractive force between  two point-like masses being given by the famous inverse square law\cite{Newton}

\begin{equation}
F=G\frac{m_1m_2}{r^2}.
\label{eqn:Inverse}\end{equation}
Let us consider  galaxies  interacting only via this force law, and so pressure-less.  To model the Hubble expansion we take this collection of  galaxies to have large-scale mass-density homogeneity and expanding  in the Hubble manner, i.e. with  a radial speed   $v(r,t)$ proportional to the distance $r$ from any particular observer. Only this Hubble law is consistent with a centre-less expansion. 
Then the well-known energy equation for any particular galaxy of mass $m$ distance $r$ from the observer is 
\begin{equation}
\frac{1}{2}m v(r,t)^2-G\frac{mM(r,t)}{r}=E,
\label{eqn:energy}\end{equation}
where $M(r,t)$ is the total mass enclosed in the sphere of radius  $r$ at time $t$.  This simply express the galactic energy $E$ as  the sum of a kinetic energy and a gravitational potential energy.  We shall include in $M(r,t)$ the mass equivalent of any other energies present, such as EM radiation, neutrino energies, and the putative dark matter and dark energies.  $M(r,t)$ is trivially given by
\begin{equation}
M(r,t)=\frac{4}{3}\pi r^3 G\rho(t),
\end{equation}
where $\rho(t)$ is the effective matter density at time $t$. For the critical case of $E=0$, (\ref{eqn:energy}) gives
\begin{equation}
v(r,t)=H(t)r   \mbox{\ \ \ \ \   where \ \ \ \ \ } H(t)=\sqrt{\frac{8}{3}\pi G \rho(t)}.
\label{eqn:HubbleNewton}\end{equation}
which gives the well-known relationship between  the Hubble expansion `constant' $H(t)$ and the effective matter density $\rho(t)$.  One of the fundamental problems in cosmology  has been that the current-epoch observed value of $\rho(t)$  is only 5\% of that predicted  from (\ref{eqn:HubbleNewton}) using  the observed value of $H(t)$, as discussed later.  In any case the above model claims that the expansion  of the universe, as given by $v(r,t)$,  is determined solely by the Newtonian gravitational force between  the effective matter content of the universe.  To put this in the form of the current  $\Lambda$CDM model
we first introduce the scale factor $a(t)=r(t)/r(t_0)$ for some fixed $t_0$, then   (\ref{eqn:HubbleNewton}) becomes
\begin{equation}
\dot{a}(t)^2={\frac{8}{3}\pi G \rho(t)}a(t)^2 
\label{eqn:Friedmann}\end{equation}
which is the Friedmann equation in the case of  a flat-space universe. To determine the time evolution of $a(t)$ we need to only specify the time-evolution of $\rho(t)$.
The  validity of (\ref{eqn:Friedmann}) is taken for granted in the analysis of the supernovae magnitude-redshift data. In order to fit that data  it  was found \cite{S1,S2} that an acceptable fit could only be obtained if $\rho(t)$ was taken to have the form
\begin{equation}
\rho(t)= \Lambda+\frac{\rho_m}{a(t)^3}
\label{eqn:fit}\end{equation}
corresponding to an effective matter-density $\Lambda$ that remains constant as the universe expands, and which is variously known as `dark energy' or the `cosmological constant', and a component which diminished like  $1/a^3$, as would happen for normal matter. However the best-fit value for the constant $\rho_m$ exceeds the known actual matter density by a factor of  5 or more, and so the remainder was interpreted as `dark matter'.
Then $\rho_m=\rho_{DM} + \rho_b$ is broken down into two components, $\rho_{DM}$ and $\rho_b$, being the supposed dark matter density and the normal observed (baryonic ) matter density, respectively. 
An EM radiation term, which diminishes as $1/a^4$, could be included, but only plays a role in the very early epoch.  We thus obtain the $\Lambda$CDM model dynamics:
\begin{equation}
\dot{a}(t)^2=\frac{8\pi G}{3}\left(\Lambda +\frac{\rho_{m}}{a^3}\right)a(t)
\label{eqn:scaleeqn}\end{equation}
  Using this Newtonian cosmological model the best-fit values for $\Lambda$ and $\rho_m$ may be determined from the supernovae data.
The parameters $\Omega_\Lambda$ etc   are defined by the fractions
\begin{equation}
\Omega_\Lambda=\Lambda/(\Lambda+\rho_{DM} + \rho_b),
\label{eqn:omegaa}\end{equation}
and so on. Because the $\Lambda CDM$ model in  (\ref{eqn:scaleeqn}) is based upon Newtonian gravity the expansion rate of the universe in this model is determined by its energy content, as expressed by $\Lambda, \rho_{DM}$ and $ \rho_m$, and in earlier epochs $\rho_r$ - the radiation density parameter.  This means that a universe without energy content cannot expand. The reason for this is that  in Newtonian gravity expansion is defined by the separation of matter - there is no notion of space itself expanding. Indeed in Newtonian gravity space has no observational or dynamical  properties - it is a totally unchanging and inert entity.  

We now briefly review a  theory for a dynamical space that has its own dynamical time evolution, which only in part is  determined by the presence of matter.  Generalising the Schr\"{o}dinger equation to encompass this dynamical space we obtain a quantum theory explanation for the phenomenon of gravity.   The fundamental dynamical equation for this 3-space (see next section)  leads to the time evolution equation for the universe:
\begin{equation}
\ddot {a}(t)
=-\frac{4\pi G}{3}\frac{\rho_{m}}{a^3}a(t)
\label{eqn:space}\end{equation}
in the case of normal matter, but with extra terms shown in (\ref{eqn:Reqn}).
This equation  gives  an expanding universe even the absence of matter/energy, in which case the expansion is uniform in time.  So this expansion breaks the long-standing connection between the matter density and the Hubble constant, as in (\ref{eqn:HubbleNewton}), and which has been so problematic.  As discussed later the supernovae data actually shows that the universe expansion is very close to being uniform in time, contrary to misleading claims of an accelerating universe.   The key problem of the  $\Lambda CDM$ model is that it does not have the observed uniformly expanding universe solution, unless, and approximately, the values of $\Omega_\Lambda$ and  $\Omega_m$ are judiciously chosen to have the best-fit\footnote{In doing the least-squares best-fit the distance modulus is used as a measure, in keeping with its use in \cite{S1,S2} .} values of  
$\Omega_\Lambda=0.73$ and   $\Omega_m=0.27$. As $\Omega_b=0.05$, we obtain $\Omega_{DM}=0.22$ in fitting the original supernovae data - see fig \ref{fig:difference}.

Hence the whole `dark energy - dark matter' imbroglio is simply a consequence of extending Newtonian gravity far beyond its realm of confirmation, and missing new physics that is absent in early modelling of gravity. We briefly review this new physics, and in later sections  show how this physics gives a parameter-free account of cosmology without requiring dark energy and dark matter. `Dark matter' of course has a longer history than its invocation in analysing the supernovae data. Nevertheless we also show that the dynamical space gives rise to new effects that counter    these older arguments for `dark matter'.

\section{Dynamical Space}

We review here the minimal  model for a dynamical 3-space. As well as the various confirmed gravitational 
predictions, there is also an extensive set of direct detection experiments, discussed in \cite{Dynamicalspace}, with the most recent being from the analysis of NASA/JPL  doppler shift data from
spacecraft earth-flybys \cite{Cahillflyby}.
An information-theoretic approach to modelling reality  leads to an emergent structured quantum-foam `space'  which is 3-dimensional and dynamic, but where the 3-dimensionality is only approximate, in that if we ignore non-trivial topological aspects of the quantum foam, then it may be coarse-grain embedded in a 3-dimensional  geometrical manifold.  Here the space is a real existent discrete but fractal network of relationships or connectivities,  but the embedding space is purely a mathematical way of characterising the 3-dimensionality of the network.  This is illustrated by the  skeletal representation of the quantum foam in figure \ref{fig:Embedd}b - this is not necessarily local in that significant linkages can manifest between distant regions.  
\begin{figure}[t]
\hspace{20mm}\parbox{60mm}{\hspace{10mm}\includegraphics[width=55mm]{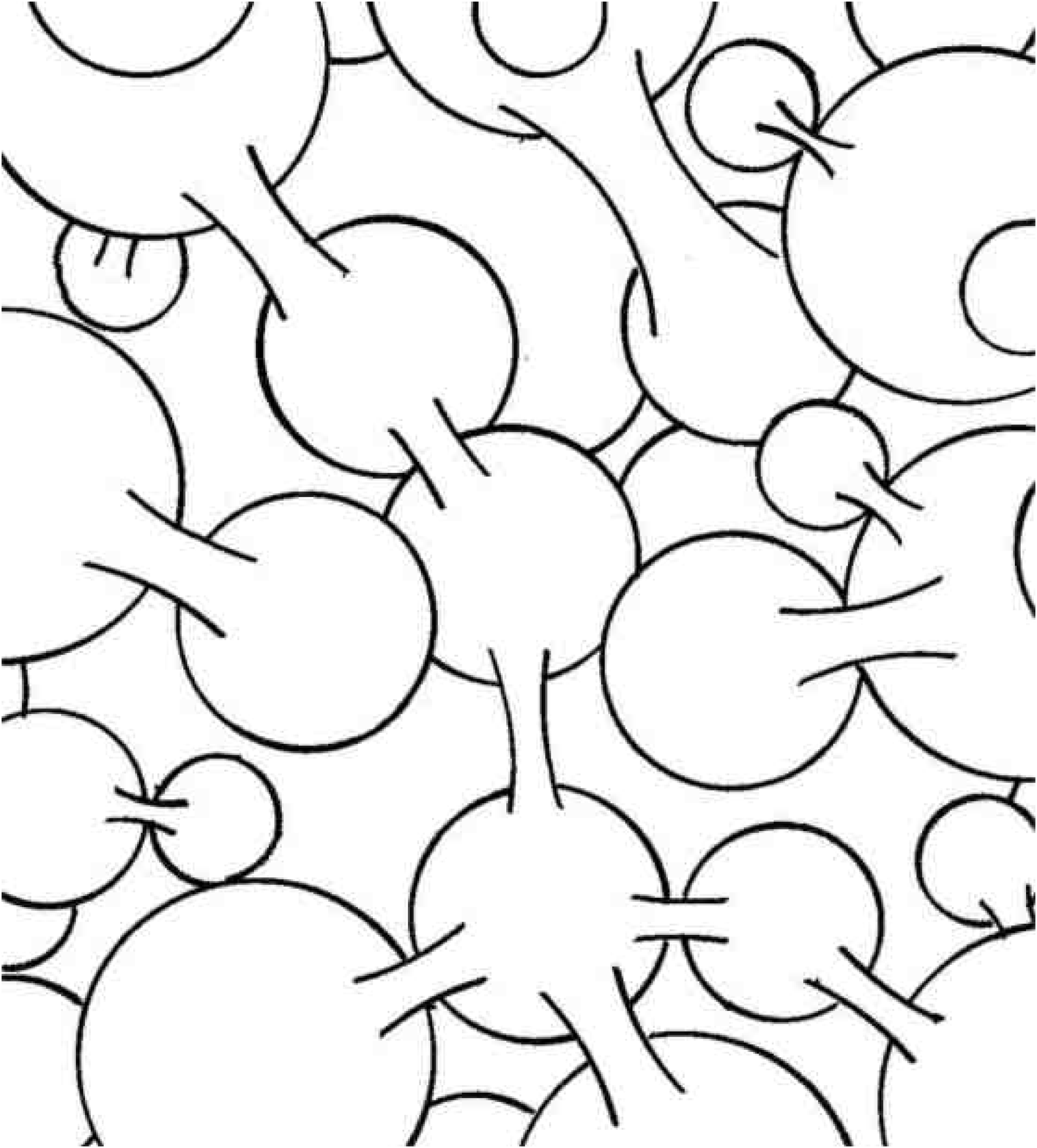}}\,
\parbox{70mm}{\vspace{0mm}\,\parbox{60mm}{\hspace{10mm}\includegraphics[width=60mm]{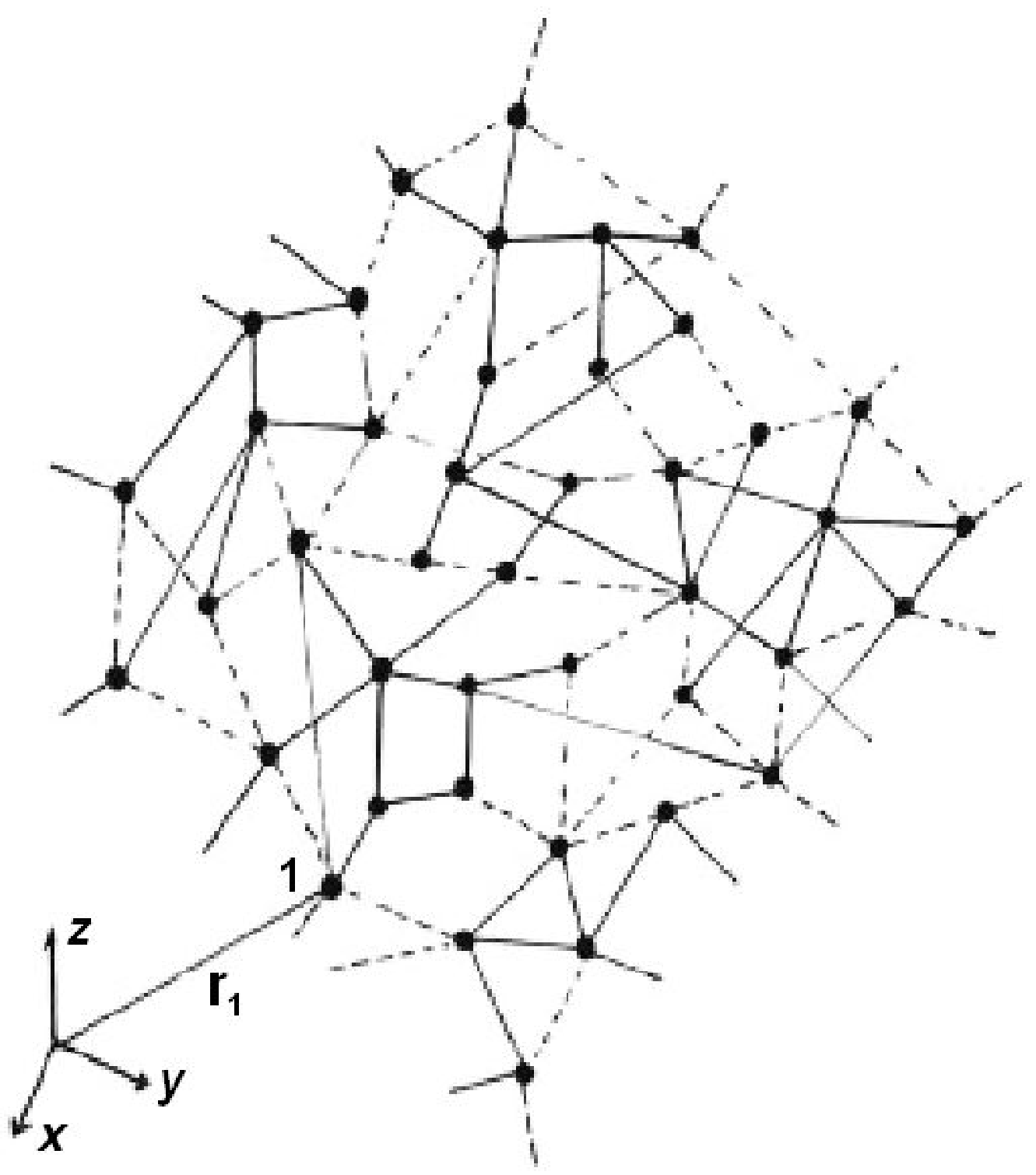}}\,}
\caption{\footnotesize{  This is an iconic representation of a quantum foam. Its skeletal structure  has its inherent approximate 3-dimensional connectivity displayed by an embedding in a mathematical   space, such as an $E^3$ or an $S^3$, as shown on the right.  The embedding space is not real - it is purely a mathematical artifact. Nevertheless this embeddability helps determine the minimal dynamics for the network, as in (\ref{eqn:E1}).   The dynamical space is not an ether model, as the embedding space does not exist.}}
\label{fig:Embedd}
\end{figure}
Embedding the network in the embedding space is very arbitrary; we could equally well rotate the embedding or use an embedding that has the network translated or translating.  These general requirements  then dictate the minimal dynamics for the actual network, at a phenomenological level.  To see this we assume  at a coarse grained level that the dynamical patterns within the network may be described by a velocity field ${\bf v}({\bf r},t)$, where ${\bf r}$ is the location of a small region in the network according to some arbitrary embedding.  The 3-space velocity field has been observed in at least 10 experiments  
\cite{Book}. 
For simplicity we assume here that the global topology of the network   is not significant for the local dynamics, and so we embed in an $E^3$, although a generalisation to an embedding in $S^3$ is straightforward and might be relevant to cosmology.  The minimal dynamics is then obtained by  writing down the sum of the only three lowest-order zero-rank tensors, of dimension $1/T^2 $, that are invariant under translation and rotation, giving
\begin{equation}
\nabla.\left(\frac{\partial {\bf v} }{\partial t}+({\bf v}.{\bf \nabla}){\bf v}\right)
+\frac{\alpha}{8}(tr D)^2 +\frac{\beta}{8}tr(D^2)=
-4\pi G\rho
\label{eqn:E1}\end{equation}
\begin{equation}D_{ij}=\frac{1}{2}\left(\frac{\partial v_i}{\partial x_j}+
\frac{\partial v_j}{\partial x_i}\right)
\label{eqn:E1b}\end{equation}
where $\rho({\bf r},t)$ is an effective matter density that may  correspond to various energy densities. The embedding space coordinates provide a coordinate system or frame of reference that is convenient to describing the velocity field, but which is not real.  

We see that there are only four possible terms, and so we need at most three possible constants to describe the dynamics of space: $G, \alpha$ and $\beta$. $G$ turns out  to be Newton's gravitational constant, and describes the rate of non-conservative flow of space into matter.  To determine the values of $\alpha$ and $\beta$ we must, at this stage, turn to experimental and observational data.  
However most data involving the dynamics of space is obtained by detecting the so-called gravitational  acceleration of matter, although increasingly light bending is giving new information.  Now the acceleration ${\bf a}$ of the dynamical patterns in space is given by the Euler or convective expression
\begin{equation}
{\bf a}({\bf r},t)= \lim_{\Delta t \rightarrow 0}\frac{{\bf v}({\bf r}+{\bf v}({\bf r},t)\Delta t,t+\Delta
t)-{\bf v}({\bf r},t)}{\Delta t} 
=\frac{\partial {\bf v}}{\partial t}+({\bf v}.\nabla ){\bf v}
\label{eqn:E3}\end{equation} 
and this appears in one of the terms in (\ref{eqn:E1}). As shown  in \cite{Schrod} and discussed later in Sect. \ref{sect:acceln} the acceleration  ${\bf g}$ of quantum matter is identical to this acceleration, apart from vorticity and relativistic effects, and so the gravitational acceleration of matter is also given by (\ref{eqn:E3}).

Outside of a spherically symmetric distribution of matter,  of total mass $M$, we find that one solution of (\ref{eqn:E1}) is the velocity in-flow field  given by
\begin{equation}
{\bf v}({\bf r})=-\hat{{\bf r}}\sqrt{\frac{2GM(1+\frac{\alpha}{2}+..)}{r}}
\label{eqn:E4}\end{equation}
but only when $\beta=-\alpha$,  for only then is the acceleration of matter, from (\ref{eqn:E3}), induced by this in-flow of the form
\begin{equation}
{\bf g}({\bf r})=-\hat{{\bf r}}\frac{GM(1+\frac{\alpha}{2}+..)}{r^2}
\label{eqn:E5}\end{equation}
 which  is Newton's Inverse Square Law of 1687 \cite{Newton}, but with an effective  mass $M(1+\frac{\alpha}{2}+..)$ that is different from the actual mass $M$.  So the success of Newton's law in the solar system, based on Kepler's analysis, informs us that  $\beta=-\alpha$ in (\ref{eqn:E1}). But we also see modifications coming from the 
$\alpha$-dependent terms.

In general because (\ref{eqn:E1}) is a scalar equation it is only applicable for vorticity-free flows $\nabla\times{\bf v}={\bf 0}$, for then we can write ${\bf v}=\nabla u$, and then (\ref{eqn:E1}) can always be solved to determine the time evolution of  $u({\bf r},t)$ given an initial form at some time  $t_0$.
The $\alpha$-dependent term in (\ref{eqn:E1})  (with now $\beta=-\alpha$) and the matter acceleration effect, now also given by (\ref{eqn:E3}),   permits   (\ref{eqn:E1})   to be written in the form
\begin{equation}
\nabla.{\bf g}=-4\pi G\rho-4\pi G \rho_{DM},
\label{eqn:E7}\end{equation}
where
\begin{equation}
\rho_{DM}({\bf r},t)\equiv\frac{\alpha}{32\pi G}( (tr D)^2-tr(D^2)),  
\label{eqn:E7b}\end{equation}
which  is an effective matter density, not necessarily non-negative,  that would be required to mimic the
 $\alpha$-dependent spatial self-interaction dynamics. The Newtonian coupling constant $G$ is included in the definition of $\rho_{DM}$ only so that its role as an effective matter density can be illustrated - the  $\alpha$ dynamics does not involves $G$.  
 Then (\ref{eqn:E7}) is the differential form for Newton's law of gravity but with an additional non-matter effective matter density.  So we label this as $\rho_{DM}$ even though no matter is involved \cite{alpha,DM}, as this effect has been shown to explain the so-called `dark matter' effect in spiral galaxies, bore hole $g$ anomalies, and the systematics of galactic black hole masses.  
 
 The spatial dynamics  is non-local.  Historically this was first noticed by Newton who called it action-at-a-distance. To see this we can write  (\ref{eqn:E1}) as an integro-differential equation
 \begin{equation}
 \frac{\partial {\bf v}}{\partial t}=-\nabla\left(\frac{{\bf v}^2}{2}\right)+G\!\!\int d^3r^\prime
 \frac{\rho_{DM}({\bf r}^\prime, t)+\rho({\bf r}^\prime, t)}{|{\bf r}-{\bf r^\prime}|^3}({\bf r}-{\bf r^\prime})
 \label{eqn:E8}\end{equation}
This shows a high degree of non-locality and non-linearity, and in particular that the behaviour of both $\rho_{DM}$ and $\rho$ manifest at a distance irrespective of the dynamics of the intervening space. This non-local behaviour is analogous to that in quantum systems and may offer a resolution to the horizon problem. As well the dynamics involving  $\rho_{DM}$ manifests at a a distance  to a scale independent of $G$, because of the $1/G$ coefficient in $\rho_{DM}$, as noted above, and so `gravitational wave' effects caused by distant activity are predicted to be much large than predicted by GR.

\begin{figure}
\hspace{35mm}\includegraphics[scale=0.3]{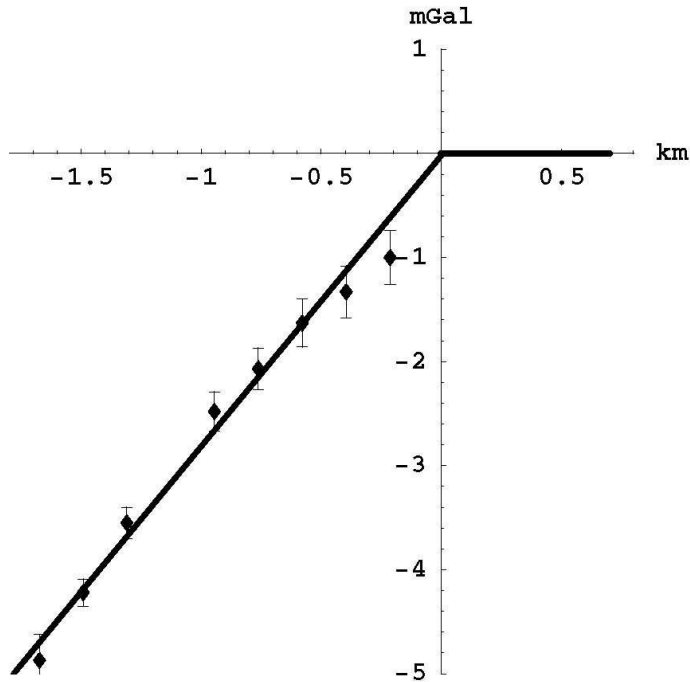}
{\caption{\small{  The data shows the gravity residuals for the Greenland Ice Shelf \cite{Ander89} Airy measurements of
the $g(r)$  profile,  defined as $\Delta g(r) = g_{Newton}-g_{observed}$, and measured in mGal (1mGal $ =10^{-3}$ cm/s$^2$)
and   plotted against depth in km. The borehole effect is that Newtonian
gravity and the new theory differ only beneath the surface, provided that the measured above-surface gravity gradient 
is used in  both theories.  This then gives the horizontal line above the surface. Using (\ref{eqn:E6}) we obtain
$\alpha^{-1}=137.9 \pm  5$ from fitting the slope of the data, as shown. The non-linearity  in the data arises from
modelling corrections for the gravity effects of the   irregular sub ice-shelf rock  topography. The ice density is 920 kg/m$^3$.}}
\label{fig:Greenland}}
\end{figure}

\begin{figure}[t]
\hspace{10mm}\,\hspace{10mm}\parbox{70mm}{\includegraphics[width=60mm,scale=0.2]{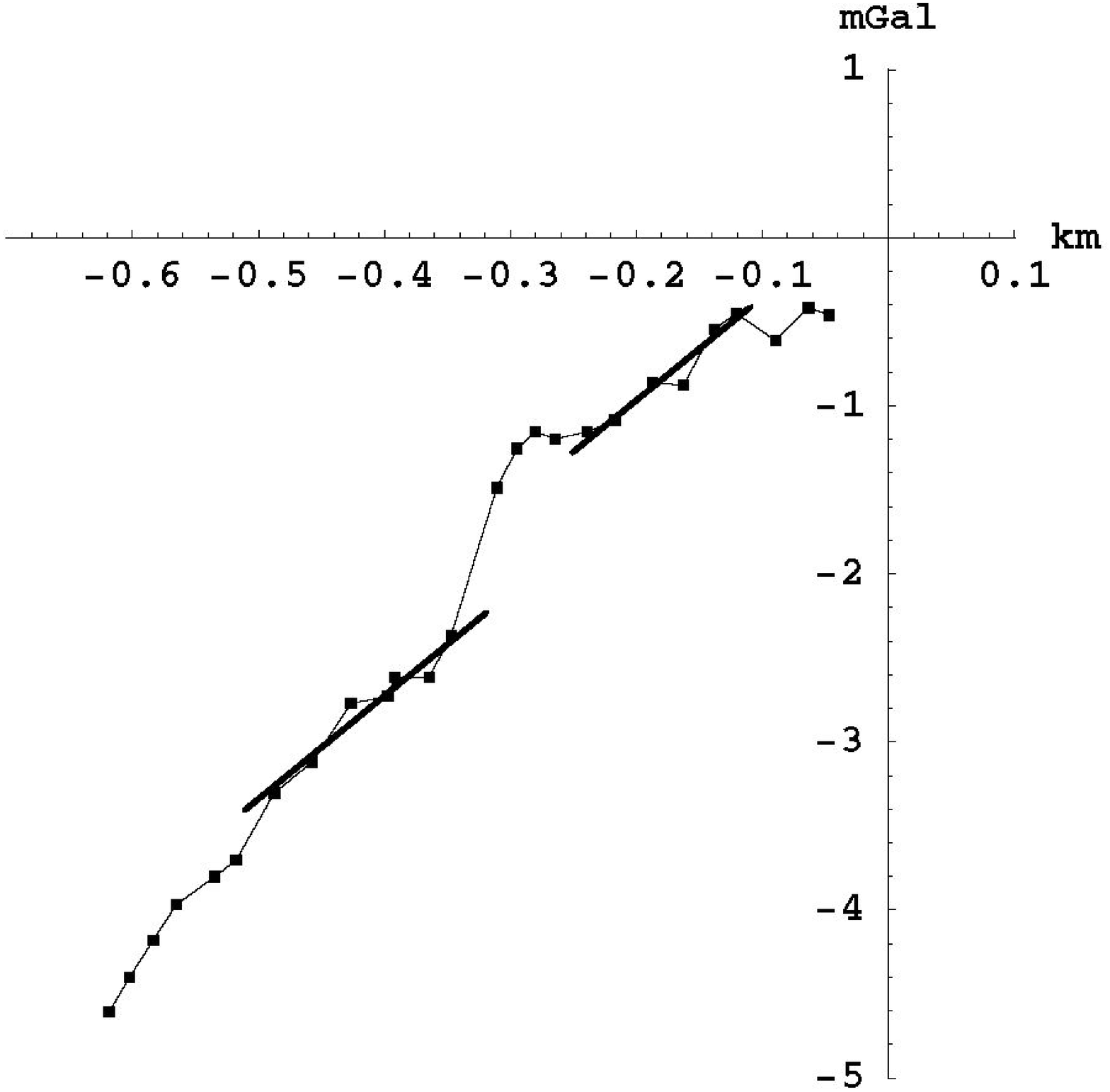}}
\parbox{70mm} {\includegraphics[width=60mm,scale=0.2]{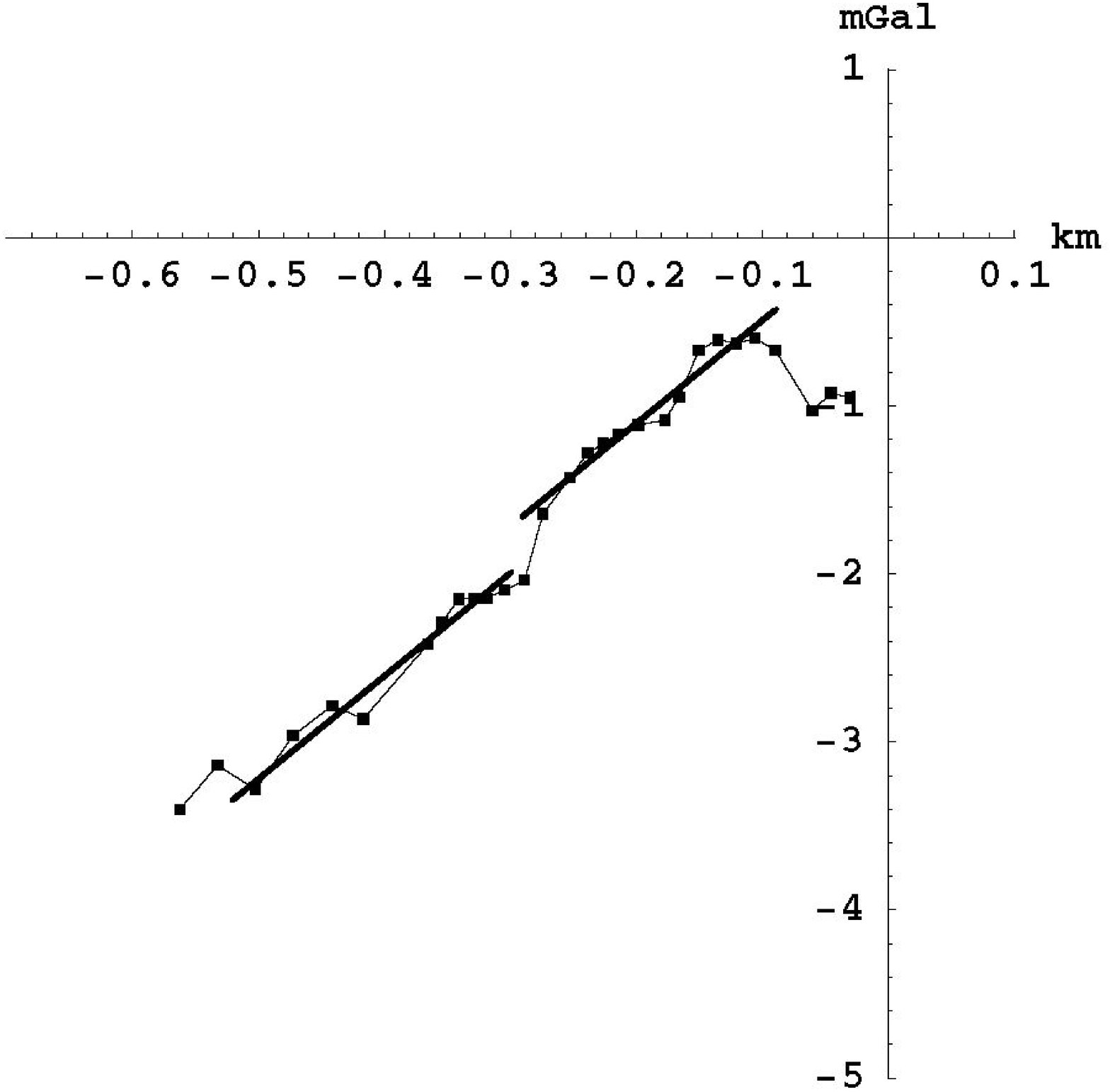}}
{\caption{\small{  Gravity residuals  $\Delta g(r) $ from two of the Nevada bore hole experiments \cite{Thomas90} that give a best fit of $\alpha^{-1}=136.8\pm 3$ on using (\ref{eqn:E6}). Some layering of the rock is evident. The rock density is 2000 kg/m$^3$ in the linear regions.}}
\label{fig:Nevada}}
\end{figure}
 
 \section{Bore Hole Anomaly: Fine Structure Constant}
 
 A recent discovery \cite{alpha, DM} has been that experimental data from the bore hole $g$ anomaly has revealed that $\alpha$ is the fine structure constant, to within experimental errors: $\alpha=e^2/\hbar c \approx 1/137.04$. This observed anomaly is that $g(r)$ does not decrease as rapidly as predicted by Newtonian gravity or GR as we descend down a bore hole.  
 Consider the case where we have a spherically symmetric matter distribution, at rest on average with respect to distant space, and that the in-flow is time-independent and radially symmetric.  Then (\ref{eqn:E1})  can now be written in the form,  with $v^\prime=dv(r)/dr$, 
 \begin{equation}
 vv^{\prime\prime}+2\frac{vv^\prime}{r} +(v^\prime)^2  =-4\pi G\rho(r)-4\pi G \rho_{DM}(v(r)), 
\label{eqn:Eradial}
\end{equation} 
where now
 \begin{equation}
\rho_{DM}(r)= \frac{\alpha}{8\pi G}\left(\frac{v^2}{2r^2}+ \frac{vv^\prime}{r}\right).
\label{eqn:E10}\end{equation}
The dynamics in (\ref{eqn:Eradial}) and (\ref{eqn:E10}) gives
 the anomaly to be
 \begin{equation}
 \Delta g=2\pi\alpha G \rho d +O(\alpha^2)
 \label{eqn:E6}\end{equation}
where $d$ is the depth and $\rho$ is the density, being that of glacial ice in the case of the Greenland Ice Shelf experiments \cite{Ander89}, or that of rock in the Nevada test site experiment \cite{Thomas90}. Clearly (\ref{eqn:E6}) permits the value of $\alpha$ to be determined from the data, giving  $\alpha=1/ (137.9 \pm 5)$ from the Greenland data, and  $\alpha=1/(136.8\pm 3)$ from the Nevada data; see Figs. \ref{fig:Greenland} and \ref{fig:Nevada}.  Note that the density  $\rho$ in (\ref{eqn:E6}) is very different for these two experiments, showing that the extracted value $\alpha$  $\approx 1/137$ is robust.

\section{Minimal and Non-Minimal Black Holes: Fine \newline Structure Constant}

\begin{figure}[t]
\hspace{15mm}\includegraphics[scale=0.4]{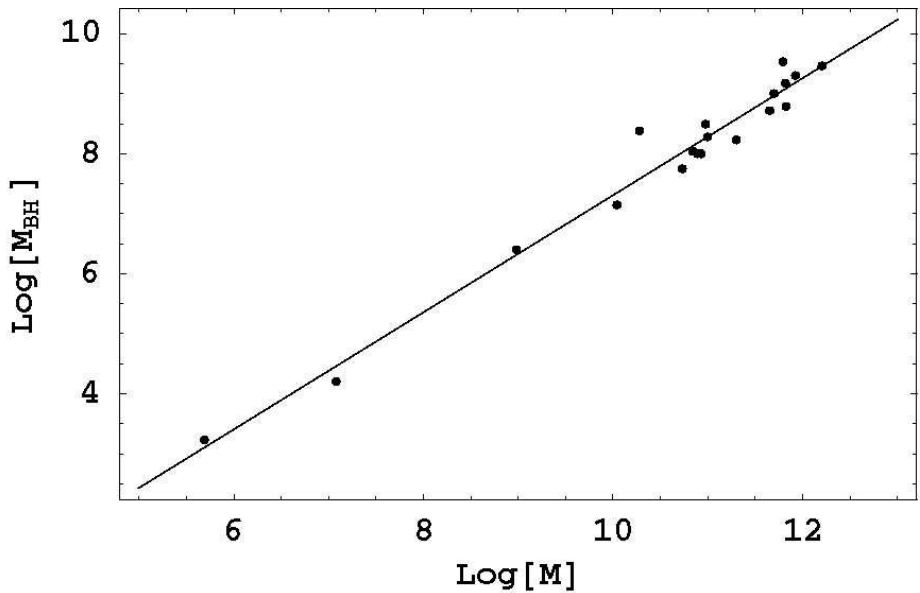}
\vspace{-4mm}
\caption{\small{The data shows $\mbox{Log}_{10}[M_{BH}]$ for the black hole masses $M_{BH}$  for
a variety of spherical matter systems, from Milky Way globular clusters to spherical galaxies, with masses $M$, plotted against 
$\mbox{Log}_{10}[M]$, in solar masses $M_0$.  The straight line is the prediction from (\ref{eqn:bhmasses}) with $\alpha=1/137$. See \cite{newBH} for references to the data. 
  \label{fig:blackholes}}}
\end{figure}

Eqn.(\ref{eqn:Eradial}) with $\rho=0$ has  exact analytic `black hole' solutions, given by (\ref{eqn:vexactb}) without the $1/r$  term.  There are two classes of black hole solutions - they are distinguished by how they relate to the surrounding matter. The class of minimal black holes is completely induced by the  surrounding distribution of matter.   For a spherically symmetric distribution of matter we find by iterating (\ref{eqn:Eradial}) and then from  (\ref{eqn:E10}) that the total effective black hole  mass is
\begin{equation}
M_{BH}=M_{DM} = 4\pi\int_0^\infty r^2\rho_{DM}(r)dr = \frac{\alpha}{2}M+O(\alpha^2)
\label{eqn:bhmasses}\end{equation}
This solution is applicable to the black holes at the centre of spherical star systems, where we identify $M_{DM}$ as $M_{BH}$.   For these black holes the acceleration $g$ outside of the matter decreases as $1/r^2$. So far black holes in 19  spherical star systems have been detected and together their masses are plotted in 
figure \ref{fig:blackholes} and compared with (\ref{eqn:bhmasses}), giving again $\alpha=1/137$ \cite{galaxies,newBH}.   These solutions are called `black holes' because they posses an event horizon that forbids the escape of EM radiation and matter, but that they are very different from the putative `black holes' of GR. Clearly GR cannot predict the mass relation in (\ref{eqn:bhmasses}) as the GR dynamics does not involve $\alpha$.  The second class of black hole solutions is called non-minimal.  These come into existence before subsequently attracting matter.  These  black holes may be primordial in that they formed directly as a consequence of the big bang before stars and galaxies, and indeed may have played a critical role in the precocious formation of  galaxies.  These black holes are responsible for both the rapid in-fall of matter to form rotating spiral galaxies, and also for  non-Keplerian rotation characteristics of these galaxies, as discussed next.  It is significant that the bore hole, black hole and (next) the spiral galaxy rotation effects are all caused by the non-local dynamics from the $\alpha$-dynamics - and so are indicative of the non-local quantum effects of the quantum cosmology.

\section{Spiral Galaxy Rotation Anomaly: Fine Structure Constant}

\begin{figure}[t]
\hspace{30mm}\includegraphics[scale=1.2]{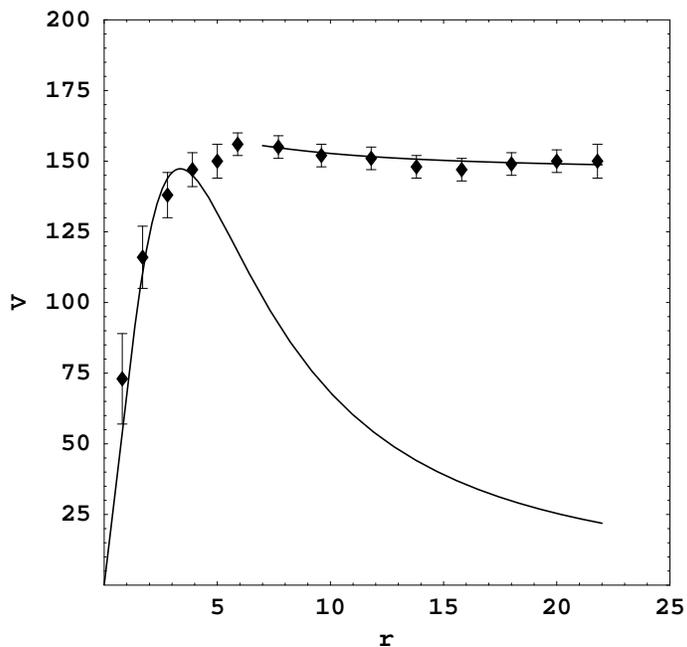}
\vspace{-4mm}
\caption{\small {Data shows the non-Keplerian rotation-speed curve $v_O$ for the spiral galaxy NGC 3198 in km/s plotted
against radius in kpc/h. Lower curve is the rotation curve from the Newtonian theory  for an
exponential disk, which decreases asymptotically like $1/\sqrt{r}$. The upper curve shows the asymptotic form from
(\ref{eqn:vorbital}), with the decrease $\sim 1/r$ determined by the small value of $\alpha$.  This asymptotic form is caused by
the primordial black holes at the centres of spiral galaxies, and which play a critical role in their formation. The
spiral structure is caused by the rapid in-fall towards these primordial black holes.}
\label{fig:NGC3198}}\end{figure}

The black hole solutions of   (\ref{eqn:Eradial}) give a direct explanation for the spiral galaxy rotation anomaly.    For a non-spherical system numerical solutions of (\ref{eqn:E1}) are required, but sufficiently far from the centre we find an exact non-perturbative two-parameter class of analytic solutions
\begin{equation}
v(r) = K\left(\frac{1}{r}+\frac{1}{R_s}\left(\frac{R_s}{r}  \right)^{\displaystyle{\frac{\alpha}{2}}}  \right)^{1/2}
\label{eqn:vexactb}\end{equation}
where $K$ and $R_s$ are arbitrary constants in the $\rho=0$ region, but whose values are determined by matching to the solution in the matter region. Here $R_s$ characterises the length scale of the non-perturbative part of this expression,  and $K$ depends on $\alpha$, $G$ and details of the matter distribution.   From (\ref{eqn:E5})  and (\ref{eqn:vexactb}) we obtain a replacement for  the Newtonian  `inverse square law' ,
\begin{equation}
g(r)=\frac{K^2}{2} \left( \frac{1}{r^2}+\frac{\alpha}{2rR_s}\left(\frac{R_s}{r}\right)
^{\displaystyle{\frac{\alpha}{2}}} 
\right),
\label{eqn:gNewl}\end{equation}
in the asymptotic limit.     The non-Newtonian part of this acceleration is caused by presence of a primordial `black hole' at the centre of the galaxy, about which the galaxy formed: in general the `black holes' from   (\ref{eqn:Eradial}) have an acceleration $g  \sim 1/r$, and very unlike the form $g \sim 1/r^2$ for the putative black holes of GR.  The centripetal acceleration  relation for circular orbits 
$v_O(r)=\sqrt{rg(r)}$  gives  a `universal rotation-speed curve'
\begin{equation}
v_O(r)=\frac{K}{2} \left( \frac{1}{r}+\frac{\alpha}{2R_s}\left(\frac{R_s}{r}\right)
^{\displaystyle{\frac{\alpha}{2}}} 
\right)^{1/2}
\label{eqn:vorbital}\end{equation}
The $\alpha$ dependent part this rotation-velocity curve  falls off extremely slowly with $r$, as is indeed observed for spiral galaxies.  This is essentially the very successful phenomenological Universal Rotation Curve for spiral galaxies \cite{URC}, but with, essentially, $\alpha \rightarrow 0$ asymptotically and the $1/r$ Keplerian term replaced by that appropriate to the in-flow into a disk of stars.  An example is shown in figure \ref{fig:NGC3198}. It was the inability of the  Newtonian  and Einsteinian gravity theories to explain these observations that led to the  notion of `dark matter'.   Note that in the absence of the $\alpha$-dynamics, the rotation-speed curve reduces to the Keplerian form.  Nevertheless  it is not clear if the  form in (\ref {eqn:vorbital}) could be used to determine the value of $\alpha$ from the extensive data set of spiral galaxy rotation curves because of observational errors and intrinsic non-systematic variations  in individual galaxies, unlike the data from bore holes and black holes which give independent but consistent determinations for the value of $\alpha$. We see that the 3-space  dynamics  (\ref{eqn:E1}) gives a unified account of both the `dark matter' problem and the properties of `black holes'.

\section{Generalised  Maxwell Equations: Gravitational \newline Lensing}

 We must   generalise the Maxwell equations so that the electric and magnetic  fields are excitations within the dynamical 3-space, and not of the embedding space.  The minimal form in the absence of charges and currents is 
 \begin{eqnarray}
\displaystyle{ \nabla \times {\bf E}}&=&\displaystyle{-\mu\left(\frac{\partial {\bf H}}{\partial t}+{\bf v.\nabla H}\right)},
 \mbox{\ \ \ }\displaystyle{\nabla.{\bf E}={\bf 0}},  \nonumber \\
 \displaystyle{ \nabla \times {\bf H}}&=&\displaystyle{\epsilon\left(\frac{\partial {\bf E}}{\partial t}+{\bf v.\nabla E}\right)},
\mbox{\ \ \  }\displaystyle{\nabla.{\bf H}={\bf 0}}\label{eqn:E18}\end{eqnarray}
which was first suggested by Hertz in 1890  \cite{Hertz}, but with ${\bf v}$ then being only a constant vector field. As easily determined  the speed of EM radiation is now $c=1/\sqrt{\mu\epsilon}$ with respect to the 3-space.
To see this we  find plane wave solutions for (\ref{eqn:E18}):
\begin{equation}
{\bf E}({\bf r},t)={\bf E}_0e^{i({\bf k}.{\bf r}-\omega t)} \mbox{\ \ \ \  } {\bf H}({\bf r},t)={\bf H}_0e^{i({\bf k}.{\bf r}-\omega t)}
\label{eqn:pw}\end{equation}
with
\begin{equation}
\omega({\bf k},{\bf v})=c|\vec{{\bf k}}| +{\bf v}.{\bf k} \mbox{ \ \ \  where \ \ \  } c=1/\sqrt{\mu\epsilon}
\label{eqn:omega}\end{equation}
Then the EM group velocity is
\begin{equation}
{\bf v}_{EM}=\vec{\nabla}_k\omega({\bf k},{\bf v})=c\hat{\bf k}+{\bf v}
\label{eqn:groupv}\end{equation}
So the velocity of EM radiation ${\bf v}_{EM}$ has magnitude  $c$ only with respect to the space, and in general not with respect to the observer if the observer is moving through space.  

The time-dependent and inhomogeneous  velocity field causes the refraction of EM radiation. This can be computed by using the Fermat least-time approximation. Then the EM ray paths  ${\bf r}(t)$ are determined by minimising  the elapsed travel time:
\begin{equation}
\tau=\int_{s_i}^{s_f}\frac{ds\displaystyle{|\frac{d{\bf r}}{ds}|}}{|c\hat{{\bf v}}_R(s)+{\bf v}(\bf{r}(s),t(s)|}
\mbox{ \ \ with \ \ }
{\bf v}_R=\left(  \frac{d{\bf r}}{dt}-{\bf v}(\bf{r}(t),t)\right)
\label{eqn:lighttime}\end{equation}
by varying both ${\bf r}(s)$ and $t(s)$, finally giving ${\bf r}(t)$. Here $s$ is a path parameter, and ${\bf v}_R$ is a 3-space tangent vector for the path.

In particular the in-flow in (\ref{eqn:E4}) causes a refraction effect of light passing close to the sun, with the angle of deflection given by
\begin{equation}
\delta=2\frac{v^2}{c^2}=\frac{4GM(1+\frac{\alpha}{2}+..)}{c^2d}
\label{eqn:E19}\end{equation}
where $v$ is the in-flow speed at distance $d$  and $d$ is the impact parameter, here the radius of the sun. Hence the  observed deflection of $8.4\times10^{-6}$ radians is actually a measure of the in-flow speed at the sun's surface, and that gives $v=615$km/s, in agreement with   the numerical value computed for $v$ at the surface of the sun from (\ref{eqn:E4}).

These generalised Maxwell equations also predict gravitational lensing produced by the large in-flows, in (\ref{eqn:vexactb}), that are the  new `black holes' in galaxies.  Until now these anomalously large lensings have been also attributed,  using GR, to the presence of `dark matter'.  One example is reported in \cite{DMGalaxies} and another in \cite{JeeDMRing} which is re-analaysed without requiring dark matter in \cite {DMRing}.

\section{Generalised  Schr\"{o}dinger Equation: Emergent \newline Gravity and Equivalence Principle \label{sect:acceln}}

A  generalisation of the  Schr\"{o}dinger equation is also required   \cite{Schrod}:
\begin{equation}
i\hbar\frac{\partial  \psi({\bf r},t)}{\partial t}=H(t)\psi({\bf r},t),
\label{eqn:equiv7}\end{equation}
where the free-fall hamiltonian is uniquely
\begin{equation}
H(t)=-i\hbar\left({\bf
v}.\nabla+\frac{1}{2}\nabla.{\bf v}\right)-\frac{\hbar^2}{2m}\nabla^2
\label{eqn:equiv8}\end{equation}
This follows from  the wave function being attached to the dynamical 3-space, and not to the embedding space, and that $H(t)$ be hermitian. We can compute the acceleration of a localised wave packet  using the Ehrenfest method \cite{Schrod}, and we obtain
\begin{equation}{\bf g}\equiv\frac{d^2}{dt^2}\left(\psi(t),{\bf r}\psi(t)\right)  
=\frac{\partial{\bf v}}{\partial t}+({\bf v}.\nabla){\bf v}+
(\nabla\times{\bf v})\times{\bf v}_R+...
\label{eqn:E11}\end{equation}
where ${\bf v}_R={\bf v}_0-{\bf v}$  is the velocity of the wave packet relative to the local space, as ${\bf v}_0$ is  the velocity relative to the embedding space. The vorticity term  causes rotation of the wave packet. For this to occur (\ref{eqn:E1}) must be generalised to the case of non-zero vorticity \cite{Book}. This vorticity effect explains the Lense-Thirring effect, and such vorticity  is being detected by the Gravity Probe B satellite gyroscope experiment \cite{GPB}. We see, as promised, that this quantum-matter acceleration is equal to that of the 3-space itself, as in (\ref{eqn:E3}). This is the first derivation of the phenomenon of gravity from a deeper theory: gravity is a quantum effect - namely the refraction of quantum waves by the internal differential motion of the substructure  patterns to 3-space itself. Note that the equivalence principle has now been explained, as this `gravitational' acceleration is independent of the mass $m$ of the quantum system. 

\section{Generalised  Dirac Equation:  Relativistic Effects in 3-Space}
An analogous generalisation of the Dirac equation is also necessary giving the coupling of the spinor to the actual dynamical 
3-space, and again not to the embedding space as has been the case up until now: 
\begin{equation}
i\hbar\frac{\partial \psi}{\partial t}=-i\hbar\left(  c{\vec{ \alpha.}}\nabla + {\bf
v}.\nabla+\frac{1}{2}\nabla.{\bf v}  \right)\psi+\beta m c^2\psi
\label{eqn:12}\end{equation}
where $\vec{\alpha}$ and $\beta$ are the usual Dirac matrices. Repeating the analysis in (\ref{eqn:E11}) for the 3-space-induced acceleration we obtain
\begin{equation}\label{eqn:E12}
{\bf g}=\displaystyle{\frac{\partial {\bf v}}{\partial t}}+({\bf v}.{\bf \nabla}){\bf
v}+({\bf \nabla}\times{\bf v})\times{\bf v}_R-\frac{{\bf
v}_R}{1-\displaystyle{\frac{{\bf v}_R^2}{c^2}}}
\frac{1}{2}\frac{d}{dt}\left(\frac{{\bf v}_R^2}{c^2}\right)+...
\label{eqn:E13a}\end{equation}
which generalises  (\ref{eqn:E11}) by having a term which limits the speed of the wave packet relative to 3-space, $|{\bf v}_R|$, to be $<\!c$. This equation specifies the trajectory of a spinor wave packet in the dynamical 3-space.  The last term causes elliptical orbits
 to precess - for circular orbits $|{\bf v}_R|$ is independent of time.

\section{Deriving the Spacetime Geodesic Formalism: \newline Local Poincar\'{e} Symmetry\label{section:spacetime}}

 We find that (\ref{eqn:E12}) may be also obtained by extremising the time-dilated elapsed time 
\begin{equation}
\tau[{\bf r}_0]=\int dt \left(1-\frac{{\bf v}_R^2}{c^2}\right)^{1/2}
\label{eqn:E13}\end{equation}  
with respect to the wave-packet trajectory ${\bf r}_0(t)$ \cite{Book}. This happens because of the Fermat least-time effect for waves: only along the minimal time trajectory do the quantum waves  remain in phase under small variations of the path. This again emphasises  that gravity is a quantum matter wave  effect.   We now introduce an effective  spacetime mathematical construct according to the metric
\begin{equation}
ds^2=dt^2 -(d{\bf r}-{\bf v}({\bf r},t)dt)^2/c^2 
=g_{\mu\nu}dx^{\mu}dx^\nu
\label{eqn:E14}\end{equation}
which is of the Panlev\'{e}-Gullstrand class of metrics \cite{Panleve,Gullstrand}. Then we have a  Local Poinacr\'{e} Symmetry, namely the transformations that leave  $ds^2$ locally invariant under a change of coordinates.   As well wave effects from (\ref{eqn:E1}) cause `ripples' in this induced spacetime, giving a different account of gravitational waves.  
The elapsed time in (\ref{eqn:E13}) may then be written as 
\begin{equation}
\tau=\int dt\sqrt{g_{\mu\nu}\frac{dx^{\mu}}{dt}\frac{dx^{\nu}}{dt}}.
\label{eqn:E14b}\end{equation}
The minimisation of  (\ref{eqn:E14b}) leads to the geodesics of the spacetime, which are thus equivalent to the trajectories from (\ref{eqn:E13}), namely (\ref{eqn:E13a}). 
We may introduce the  standard differential geometry curvature tensor for the induced  spacetime
\begin{equation}
R^\rho_{\mu\sigma\nu}=\Gamma^\rho_{\mu\nu,\sigma}-\Gamma^\rho_{\mu\sigma,\nu}+
\Gamma^\rho_{\alpha\sigma}\Gamma^\alpha_{\mu\nu}-\Gamma^\rho_{\alpha\nu}\Gamma^\alpha_{\mu\sigma},
\label{eqn:curvature}\end{equation}
where $\Gamma^\alpha_{\mu\sigma}$ is the affine connection for the metric in (\ref{eqn:E14})
\begin{equation}
\Gamma^\alpha_{\mu\sigma}=\frac{1}{2} g^{\alpha\nu}\left(\frac{\partial g_{\nu\mu}}{\partial x^\sigma}+
\frac{\partial g_{\nu\sigma}}{\partial x^\mu}-\frac{\partial g_{\mu\sigma}}{\partial x^\nu} \right).
\label{eqn:affine}\end{equation}
with  $g^{\mu\nu}$  the matrix inverse of  $g_{\mu\nu}$. 
In this formalism the trajectories of quantum-matter wave-packet test objects are determined by
\begin{equation}
\frac{d^2x^\lambda}{d\tau^2}+\Gamma^\lambda_{\mu\nu}\frac{dx^\mu}{d\tau}\frac{dx^\nu}{d\tau}=0,
\label{eqn:33}\end{equation}
as this is equivalent to (\ref{eqn:E12}).  In the standard treatment of GR  the geodesic for classical matter in  (\ref{eqn:33}) is a definition, and has no explanation.  Here we see that it is finally derived, but as a quantum matter effect.
Hence by coupling the Dirac spinor dynamics to the dynamical 3-space  we derive the geodesic formalism of General Relativity as a quantum effect, but without reference to the Hilbert-Einstein equations for the induced metric.  Indeed in general the metric of  this induced spacetime will not satisfy  these equations as the dynamical space involves the $\alpha$-dependent  dynamics, and $\alpha$ is missing from GR.  
We can also define the Ricci curvature scalar 
\begin{equation} R=g^{\mu\nu}R_{\mu\nu}\label{eqn:Ricci}\end{equation} 
where $R_{\mu\nu}=R^\alpha_{\mu\alpha\nu}$.  In general the induced spacetime in (\ref{eqn:E14}) has a non-zero Ricci scalar - it is a curved spacetime. We shall compute the Ricci  scalar for the expanding 3-space solution below.

We can also derive the Schwarzschild metric without reference to GR.  To do this we merely have to identify the induced spacetime metric corresponding to the in-flow in (\ref{eqn:E4}) outside of a spherical matter system, such as the earth.  Then (\ref{eqn:E14})  becomes
 \begin{equation}
ds^2=dt^{ 2}-\frac{1}{c^2}(dr+\sqrt{\frac{2GM(1+\frac{\alpha}{2}+..)}{r}}dt)^2
-\frac{r^2}{c^2}(d\theta^{ 2}+\sin^2(\theta)d\phi^2)
\label{eqn:GRE15}\end{equation}
 Making the change of variables $t\rightarrow t^\prime$ and
$\bf{r}\rightarrow {\bf r}^\prime= {\bf r}$ with
\begin{eqnarray}
t^\prime=&& t-
\frac{2}{c}\sqrt{\frac{2 GM(1{+}\frac{\alpha}{2}{+}\dots)r}{c^2}}+  \nonumber \\
&&\frac{4\ GM(1{+}\frac{\alpha}{2}{+}\dots)}{c^3}\,\mbox{tanh}^{-1}\sqrt{\frac{2 GM(1{+}\frac{\alpha}{2}{+}\dots)}{c^2r}}
\label{eqn:GRE16}\end{eqnarray}
this becomes (and now dropping the prime notation)
\begin{eqnarray}
ds^2&=&\left(1-\frac{2GM(1+\frac{\alpha}{2}+..)}{c^2r}\right)dt^{ 2} 
-\frac{1}{c^2}r^{ 2}(d\theta^2+\sin^2(\theta)d\phi^2) \nonumber \\
&&-\frac{dr^{ 2}}{c^2\left(1-{\displaystyle\frac{
2GM(1+\frac{\alpha}{2}+..)}{ c^2r}}\right)}.
\label{eqn:GRE17}\end{eqnarray}
which is  one form of the the Schwarzschild metric but with the $\alpha$-dynamics induced effective mass shift. Of course this is only valid outside of the spherical matter distribution, as that is the proviso also on (\ref{eqn:E4}). 
Hence in the case of the Schwarzschild metric the dynamics missing from both the Newtonian theory of gravity and General Relativity is merely hidden in a mass redefinition, and so didn't affect the various standard tests of GR, or even of Newtonian gravity.   A non-spherical symmetry version of the Schwarzchild metric is used in modelling the Global Positioning System (GPS).

\section{Supernova and Gamma-Ray-Burst Data} 

In the next section we show that the 3-space dynamics in (\ref {eqn:E1}) has an expanding space solution.
The supernovae and gamma-ray bursts provide standard candles that enable observation  of the expansion of the universe.  To test yet further that dynamics we compare the predicted expansion against the observables, namely the magnitude-redshift data from supernovae and gamma-ray bursts.
 The supernova data set used herein and shown in Figs. \ref{fig:SN1} and \ref{fig:SN2} is available at \cite{data set}.    Quoting from  \cite{data set}  we note that Davis {\it et al.} \cite{Davis}  combined several data sets by taking  the ESSENCE data set from Table 9 of Wood--Vassey {\it et al.}  (2007) \cite{WV}, using only the supernova that passed the light-curve-fit quality criteria. They took the HST data from Table 6 of Riess {\it et al.} (2007) \cite{Riess}, using only the supernovae classified as gold.
To put these data sets on the same Hubble diagram  Davis {\it et al.} used 36 local supernovae that are in common between these two data sets. When discarding supernovae with $z<0.0233$ (due to larger uncertainties in the peculiar velocities) they found an offset of $0.037 \pm 0.021$ magnitude between the data sets, which they then corrected for by subtracting this constant from the HST data set. The dispersion in this offset was also accounted for in the uncertainties.
The HST data set had an additional 0.08 magnitude added to the distance modulus errors to allow for the intrinsic dispersion of the supernova luminosities. The value used by Wood--Vassey {\it et al.}  (2007) \cite{WV} was instead 0.10 mag. Davis {\it  et al.}  adjusted for this difference by putting the Gold supernovae on the same scale as the ESSENCE supernovae. Finally, they also added the dispersion of 0.021 magnitude introduced by the simple offset described above to the errors of the 30 supernovae in the HST data set. The final supernova data base for  the distance modulus $\mu_{obs}(z)$ is shown in Figs. \ref{fig:SN1} and \ref{fig:SN2}.  The gamma-ray-burst (GRB) data is from Schaefer \cite{GRB}.

\section{Expanding Universe from Dynamical 3-Space}

Let us now explore the expanding  3-space  from (\ref {eqn:E1}).  Critically, and unlike the FLRW-GR model, the 3-space expands even when the energy density is zero.
Suppose that  we have a radially symmetric effective density $\rho(r,t)$, modelling EM radiation, matter, cosmological constant etc,   and that we look for a radially symmetric time-dependent flow ${\bf v}({\bf r},t) =v(r,t)\hat{\bf r}$ from (\ref{eqn:E1}) (with $\beta=-\alpha$).  Then $v(r,t)$ satisfies the equation,  with $v^\prime=\displaystyle{\frac{\partial v(r,t)}{\partial r}}$,
\begin{equation}
\frac{\partial}{\partial t}\left( \displaystyle{\frac{2v}{r}}+v^\prime\right)+vv^{\prime\prime}+2\frac{vv^{\prime}}{r}+ (v^\prime)^2+\frac{\alpha}{4}\left(\frac{v^2}{r^2} +\frac{2vv^\prime}{r}\right)
=- 4\pi G \rho(r,t)  \label{eqn:radialflow}\end{equation}
Consider first the zero energy case $\rho=0$. Then we have a Hubble  solution $v(r,t)=H(t)r$, a centreless flow, determined by
\begin{equation}{\dot H}+\left(1+\frac{\alpha}{4}\right)H^2=0
\end{equation}
with ${\dot H}=\displaystyle{\frac{dH}{dt}}$.  We also introduce in the usual manner the scale factor $a(t)$ according to $H(t)=\displaystyle{\frac{1}{a}\frac{da}{dt}}$. We then obtain
the solution
\begin{equation}
H(t)=\frac{1}{(1+\frac{\alpha}{4})t}=H_0\frac{t_0}{t}; \mbox{\ \  }  a(t)=a_0\left(\frac{t}{t_0} \right)^{4/(4+\alpha)}
\label{eqn:spacexp}\end{equation}
where $H_0=H(t_0)$ and $a_0=a(t_0)$.  Note that we obtain an expanding 3-space even where the energy density is zero - this is in sharp contrast to the FLRW-GR model for the expanding universe, as shown below.

We can write the  Hubble function $H(t)$ in terms of $a(t)$ via the inverse function $t(a)$, i.e. $H(t(a))$ and finally as $H(z)$, where the redshift observed now, $t_0$, relative to the wavelengths at time $t$, is  $z=a_0/a-1$. Then we obtain
\begin{eqnarray}
H(z)={H_0}(1+z)^{1+\alpha/4}
\label{eqn:H2a}\end{eqnarray}
To test this expansion we need to predict the relationship between the cosmological observables, namely the relationship between the apparent energy-flux magnitudes and redshifts. This  involves taking account of the reduction in photon count caused by the expanding 3-space, as well as the accompanying reduction in photon energy. To that end we first
 determine the distance travelled by the light from a supernova or GRB  event before detection.  Using a choice of embedding-space coordinate system with $r=0$ at the location of a supernova/GRB event  the 
speed of light relative to this embedding space frame is $c+v(r(t),t)$, i.e. $c$ wrt the space itself, as noted above,  where $r(t)$ is the embedding-space distance from the source. Then the distance travelled by the light at time $t$ after emission at time $t_1$ is determined implicitly by
\begin{equation}
r(t)=\int_{t_1}^t dt^\prime(c+v(r(t^\prime), t^\prime),
\label{eqn:distance1}\end{equation}
which has the solution on using $v(r,t)=H(t)r$
\begin{equation}
r(t)=c a(t)\int_{t_1}^t \frac{dt^\prime}{a(t^\prime)}.
\label{eqn:distance2}\end{equation}
This distance gives directly the surface area $4\pi r(t)^2$ of  the expanding sphere  and so the decreasing photon count per unit of that surface area. However  also because of the expansion the flux of photons is reduced by the factor  $1/(1+z)$, simply because they are spaced further apart by the expansion. The photon flux is then given by
\begin{equation}
{\cal F}_P=\frac{{\cal L}_P}{4\pi r(t)^2(1+z)}
\end{equation}
where  ${\cal L}_P$ is the source photon-number luminosity. However usually the energy flux is measured, and the energy of each photon is reduced by the factor $1/(1+z)$ because of the redshift. Then the energy flux is, in terms of the source energy luminosity ${{\cal L}_E}$, 
\begin{equation}
{\cal F}_E=\frac{{\cal L}_E}{4\pi r(t)^2(1+z)^2}\equiv\frac{{\cal L}_E}{4\pi r_L(t)^2}
\end{equation}
which  defines the effective energy-flux luminosity distance $r_L(t)$.
Expressed in terms of the observable redshift $z$ this gives an energy-flux  luminosity effective distance
\begin{equation}
r_L(z)=(1+z)r(z)=c (1+z)\int_{0}^z \frac{dz^\prime}{H(z^\prime)}
\label{eqn:distance3}\end{equation}
The dimensionless `energy-flux'  luminosity effective distance is then given by
 \begin{equation}
d_L(z)=(1+z)\int_0^z \frac{H_0 dz^\prime}{H(z^\prime)}
\label{eqn:H1a}\end{equation}
 and the theory distance modulus is  defined by
\begin{equation}
\mu(z)=5\log_{10}(d_L(z))+m.
\label{eqn:H1b}\end{equation}
Because all the selected supernova have the same absolute magnitude, $m$ is a constant whose value is determined by fitting the low $z$ data. The GRB magnitudes have been adjusted to match the supernovae data \cite{GRB}.

Using the  Hubble expansion (\ref{eqn:H2a}) in (\ref{eqn:H1a}) and (\ref{eqn:H1b}) we obtain the middle curves (red) in Figs. \ref{fig:SN1} and  the \ref{fig:SN2}, yielding an excellent agreement with the supernovae and GRB data. Note that because $\alpha/4$ is so small it actually has negligible effect on these plots.  But that is only the case for the homogeneous expansion - we saw above that the $\alpha$ dynamics can result in large effects such as black holes and large spiral galaxy rotation effects when the 3-space is inhomogeneous.  Hence the dynamical 3-space gives an immediate account of the universe expansion data, and does not require the introduction of  a cosmological constant or `dark energy', but which will be nevertheless discussed next. 

\begin{figure}
\vspace{0mm}
\hspace{15mm}\includegraphics[scale=0.6]{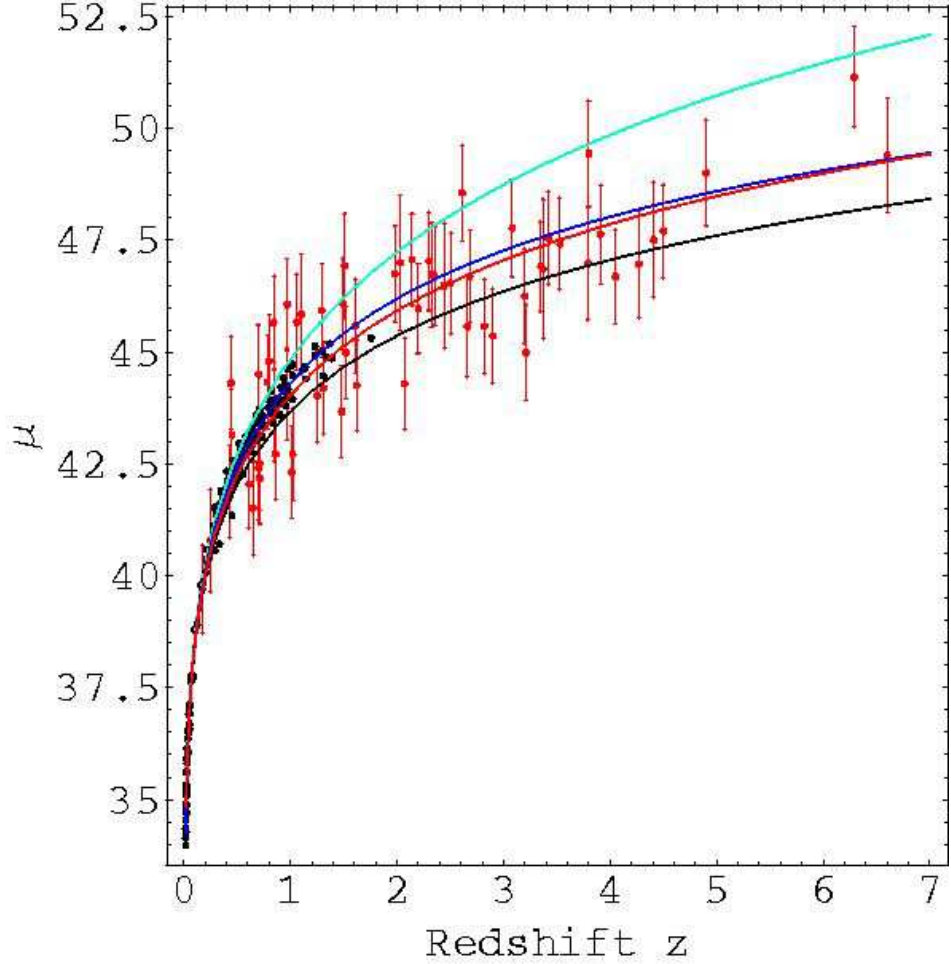}
\vspace{-5mm}\caption{\small{ Hubble diagram showing the combined supernovae data from Davis {\it et al.} \cite{Davis} using several data sets from   Riess {\it et al.} (2007)\cite{Riess} and Wood-Vassey {\it et al.}  (2007)\cite{WV} (dots without error bars for clarity - see figure \ref{fig:SN2} for error bars) and the Gamma-Ray-Bursts data (with error bars) from Schaefer \cite{GRB}.  Upper curve (green)  is `dark energy' only $\Omega_\Lambda=1$, lower curve  (black) is matter only $\Omega_m=1$. Two middle curves show best-fit of `dark energy'-`dark-matter' (blue) and dynamical 3-space prediction (red), and are essentially indistinguishable.  However the theories make very different predictions for the future. We see that the best-fit `dark energy'-`dark-matter' curve essentially converges on the uniformly-expanding parameter-free dynamical 3-space prediction. See figure \ref{fig:difference} for comparison out to $z=14$.}
\label{fig:SN1}}\end{figure}

\begin{figure}
\vspace{3.0mm}
\hspace{15mm}{\includegraphics[scale=0.6]{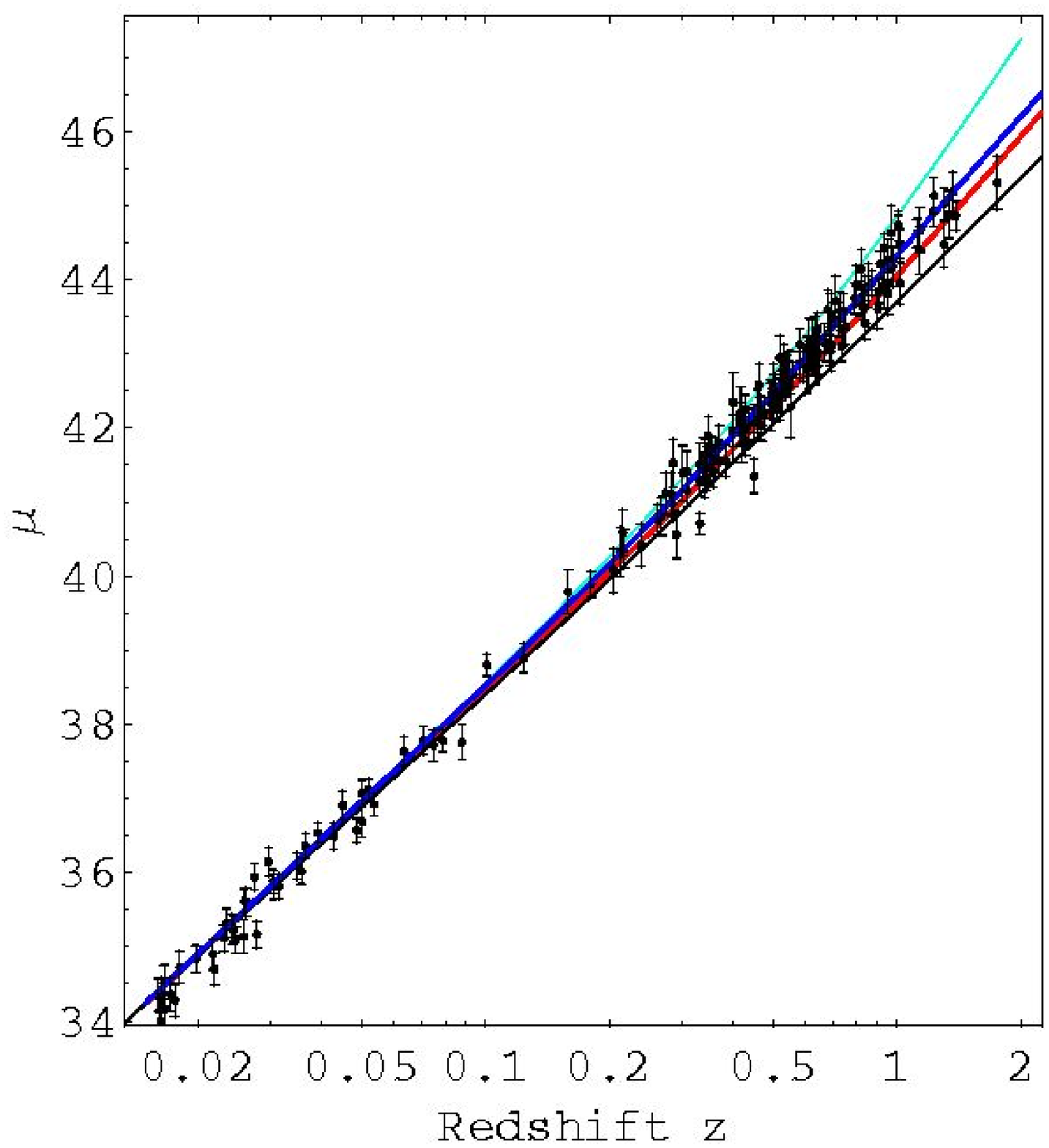}}\,
{\caption{\small{  Hubble diagram as in figure \ref{fig:SN1} but plotted logarithmically  to reveal details for  $z<2$, and without GRB data.  Upper curve (green) is `dark energy' only $\Omega_\Lambda=1$. Next curve (blue) is  best fit of `dark energy'-`dark-matter'. Lowest curve (black) is matter only $\Omega_m=1$. 2nd lowest curve (red) is  dynamical 3-space prediction. }\label{fig:SN2}}}
\end{figure}

\section{Expanding Universe - Non-Zero Energy Density Case}
When the energy density is not zero we need to take account of the dependence of $\rho(r,t)$ on the scale factor of the universe. In the usual manner we thus write
\begin{equation}
\rho(r,t)=\frac{\rho_{m}}{a(t)^3}+\frac{\rho_{r}}{a(t)^4}+\Lambda
\label{eqn:evolve}\end{equation}
for matter, EM radiation and the cosmological constant or `dark energy' $\Lambda$, respectively, where the matter and radiation is approximated by a spatially uniform (i.e independent of $r$)  equivalent matter density. We argue here that $\Lambda$ - the cosmological constant or dark energy density, like dark matter, is an unnecessary concept. We have chosen a definition for the cosmological constant $\Lambda$ so that it has the units of matter density.
Then (\ref{eqn:radialflow}) becomes for $a(t)$
\begin{eqnarray}
\frac{\ddot  a}{a}+\frac{\alpha}{4}\frac{{\dot a}^2}{a^2}
=-\frac{4\pi G}{3}\left(\frac{\rho_{m}}{a^3}+\frac{\rho_{r}}{a^4}+\Lambda \right)
\label{eqn:Reqn}\end{eqnarray}giving
\begin{equation}
{\dot a}^2=\frac{8\pi G}{3}\left(\frac{\rho_{m}}{a}+\frac{\rho_{r}}{a^2}+\Lambda a^2\right)-\frac{\alpha}{2}\int\frac{{\dot a}^2}{a}da+f
\label{eqn:R2}\end{equation}
where $f$ is an integration constant.
 In terms of ${\dot a}^2$ this has the solution
\begin{equation}{\dot a}^2\!=\!\frac{8\pi G}{3}\!\!\left(\!\frac{\rho_{m}}{(1-\frac{\alpha}{2})a}\!+\!\frac{\rho_{r}}{(1-\frac{\alpha}{4})a^2}\!+\!\frac{\Lambda a^2}{(1+\frac{\alpha}{4})}\!+\!b a^{-\alpha/2}\!\right)
\label{eqn:R3}\end{equation}
which is easily checked by substitution into (\ref{eqn:R2}), and where $b$ is the  integration constant. Finally we obtain from  (\ref{eqn:R3})
\begin{equation}
t(a)=t(a_0)+\int^a_{a_0}\frac{da}{\sqrt{\displaystyle{\frac{8\pi G}{3}}\left(\displaystyle{\frac{\rho_{m}}{a}+\frac{\rho_{r}}{a^2}}+\Lambda a^2+b a^{-\alpha/2}\right)}}
\label{eqn:R4}\end{equation} 
where  we have re-scaled the various density parameters for notational convenience.  When $\rho_m=\rho_r=\Lambda=0$, (\ref{eqn:R4})
reproduces the expansion in (\ref{eqn:spacexp}), and so the density terms in (\ref{eqn:R3}) give the modifications to  the dominant purely spatial expansion, which we have noted above  already gives an excellent account of the data. It is important to note that  (\ref{eqn:R3}) has the  $b$ term  - the constant of integration, even when $\alpha=0$, whereas the FLRW-GR dynamics demands, effectively, $b=0$. Having $b\neq 0$ simply asserts that the 3-space can expand even when the energy density is zero - an effect missing from FLRW-GR cosmology.

From (\ref{eqn:R3})  we then obtain
\begin{equation}
H(z)^2={H_0}^2(\Omega_m(1+z)^3+\Omega_r(1+z)^4 +\Omega_\Lambda+\Omega_s(1+z)^{2+\alpha/2})
\label{eqn:H2}\end{equation}
where 
\begin{equation}
H_0=\left(\frac{8\pi G}{3}(\rho_m+\rho_r+\Lambda+b)\right)^{1/2}
\label{eqn:Hubble constant}\end{equation}
\begin{equation}
\Omega_m=\rho_m/(\rho_m+\rho_r+\Lambda+b),...
 \end{equation}
and so
\begin{equation}
\Omega_m+\Omega_r+\Omega_\Lambda+\Omega_s=1.
\end{equation}

Next we discuss  the strange feature of the  FLRW-GR dynamics which requires a non-zero energy density for the universe to expand.

\begin{figure}
\hspace{-5mm}{\includegraphics[scale=0.5]{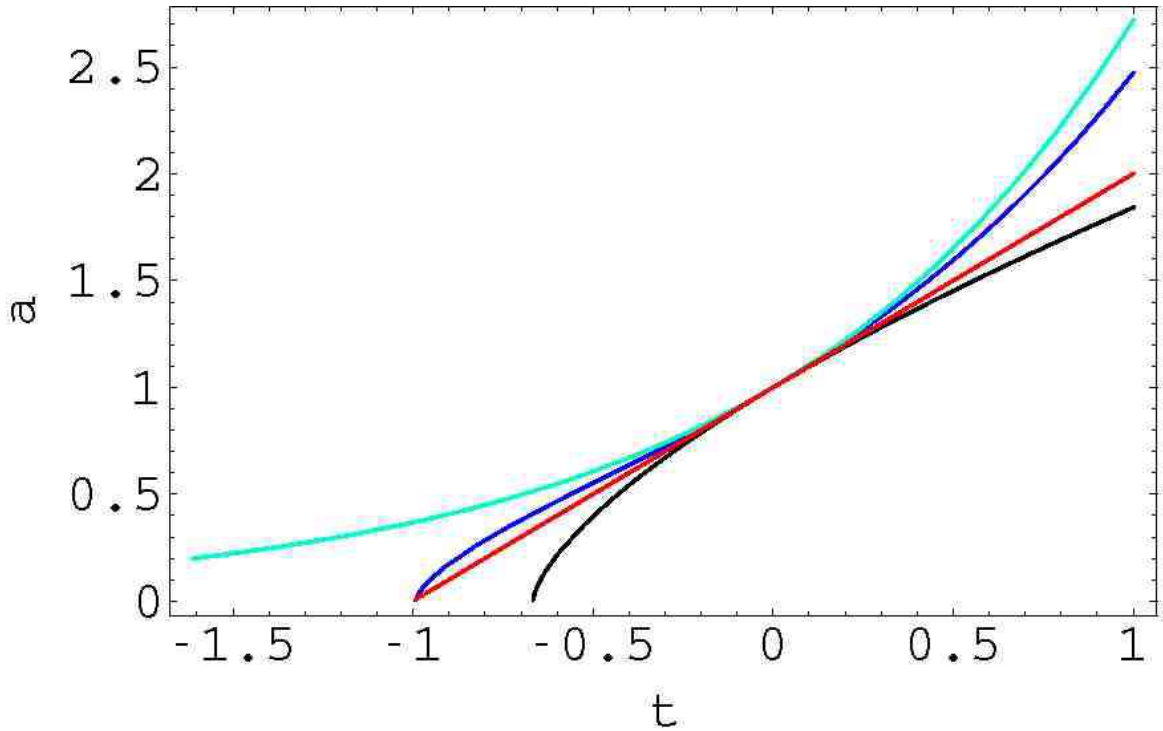}}\,
{\vspace{-4mm}\caption{\small{  Plot of the scale factor $a(t)$ vs $t$, with $t=0$ being `now' with $a(0)=1$, for the four cases discussed in the text, and corresponding to the plots in  Figs. \ref{fig:SN1} and \ref{fig:SN2}: (i) the upper  curve (green) is the `dark energy' only case, resulting in an exponential acceleration at all times, (ii) the bottom curve (black) is the matter only prediction, (iii) the 2nd highest curve (to the right of $t=0$) is the fitted  `dark energy' plus `dark-matter 'case (blue) showing a past deceleration and future exponential acceleration effect. The straight line plot  (red)  is the dynamical 3-space prediction.  We see that the best-fit `dark energy' - `dark matter' curve closely follows  the dynamical 3-space result. All plots have the same slope at $t=0$, i.e. the same value of $H_0$. 
 }\label{fig:Rtplot}}}
\end{figure}

\section{Deriving the Friedmann-Lema\^{i}tre-Robertson \newline -Walker Metric}
The induced effective spacetime metric in (\ref{eqn:E14}) is, for the Hubble expansion,
\begin{equation}
ds^2=g_{\mu\nu}dx^\mu dx^\nu=dt^2-(d{\bf r}-H(t){\bf r}dt)^2/c^2 
\label{eqn:PGmetric}\end{equation}
 The occurrence of $c$ has nothing to do with the dynamics of the 3-space - it is related to the geodesics of relativistic quantum matter, as noted above. Nevertheless changing to spatial coordinate variables   ${\bf r}^\prime$ with  ${\bf r}=a(t){\bf r}^\prime$, and with $t^\prime=t$,  we obtain
\begin{equation}
ds^2=g_{\mu\nu}dx^\mu dx^\nu=dt^{\prime2}-a(t^\prime)^2d{\bf r}^{\prime2}/c^2 
\label{eqn:Hubblemetric}\end{equation}
which is the usual   Friedmann-Lema\^{i}tre-Robertson-Walker (FLRW) metric in the case of a flat spatial section.  However this involves a deceptive choice of spacetime coordinates.  Consider the position of a galaxy located at 
${\bf r}(t)$. Then over the time interval $dt$ this galaxy moves a distance $dr={\bf v}({\bf r},t)dt=H(t){\bf r}(t)dt$. In terms of the FLRW distance however the galaxy moves through distance $d{\bf r}^\prime=d({\bf r}(t)/a(t))=(d{\bf r}(t)-H(t){\bf r}(t))/a(t)={\bf 0}$. Hence the FLRW distances involve a dynamically determined re-scaling of the spatial distance measure so that the universe does not expand in terms of these coordinates.   We now show why the FLRW cosmology model needs to invoke `dark energy' and `dark matter' to fit the observational data.The Hilbert-Einstein (HE) equations for a spacetime metric are 
\begin{equation}
G_{\mu\nu}\equiv R_{\mu\nu}-\frac{1}{2}Rg_{\mu\nu}=8\pi  G\Lambda g_{\mu\nu}+8\pi G T_{\mu\nu}
\label{eqn:HE}\end{equation}
where $G_{\mu\nu}$ is supposed to describe the dynamics of the spacetime manifold in the presence of an energy-momentum described by the  tensor $T_{\mu\nu}$. Surprisingly, in the absence of $\Lambda$ and $T_{\mu\nu}$  the HE equation, now $G_{\mu\nu}=0$, does not have an expanding universe solution for the metric in (\ref{eqn:Hubblemetric}).

The stress-energy tensor is, according to the Weyl postulate,
 \begin{equation}
T_{\mu\nu}=(\rho+p)u_\mu u_\nu+pg_{\mu\nu}
\label{eqn:stress}\end{equation}
Then with $u^\mu=(1,0,0,0)$ we obtain  for the flat spacetime in (\ref{eqn:Hubblemetric}) the well-known Friedmann equations
\begin{equation}
\frac{{\dot a}^2}{a^2}=\frac{8\pi G\Lambda}{3}+\frac{8}{3}\pi G \rho
\label{eqn:HE1}\end{equation}
\begin{equation}
\frac{{\ddot a}}{a}+\frac{{\dot a}^2}{2a^2}=4\pi G\Lambda-4\pi G p
\label{eqn:HE2}\end{equation}
These two equations constitute the dynamical equations for the current standard model of cosmology ($\Lambda$CDM).  Even in the case of zero-pressure `dust', with $p=0$, these two equations are not equivalent to (\ref{eqn:Reqn}) (with $\alpha=0$ in this section).  If $\rho=0, \Lambda=0$ and $p=0$ then these equations give the non-expanding universe $\dot a=0$, which is not the general solution to  (\ref{eqn:Reqn}) which has $\dot a$= constant, and   it is this solution which gives a parameter-free fit to the supernova/GRB redshift data. If only $p=0$ then these two equations give, first from (\ref{eqn:HE1}), and then  from  (\ref{eqn:HE1}) and  (\ref{eqn:HE2}).
\begin{equation}
\frac{{\dot a}^2}{2}-\frac{4\pi G\Lambda a^2}{3}-\frac{4\pi G \rho_m}{3a}=0
\label{eqn:p1}\end{equation}
\begin{equation}
\frac{d}{dt}\left(\frac{{\dot a}^2}{2}-\frac{4\pi G\Lambda a^2}{3}-\frac{4\pi G \rho_m}{3a}\right)=0
\label{eqn:p2}\end{equation}
Whence (\ref{eqn:p1}) requires that the integration constant from  (\ref{eqn:p2}) must be zero - this is equivalent to demanding $b=0$ in (\ref{eqn:R3}), and in the   FLRW-GR model we obtain   the well known relationship 
\begin{equation}
H_0=\left(\frac{8\pi G}{3}(\rho_m+\rho_r+\Lambda)\right)^{1/2}=\left(\frac{8\pi G\rho}{3}\right)^{1/2}
\label{eqn:FHubble constant}\end{equation}
This strict link between $H_0$ and the energy density $\rho$ has lead to the so-called `missing mass' problem: too little hadronic matter had been detected to agree with the observed value of $H_0$.   The dynamical 3-space does not have this connection between $H_0$ and $\rho$.

Hence according to the FLRW-GR dynamics the universe can only expand if at least one of $\Lambda$ or $\rho_m$ is non-zero.   This amounts to not modelling space itself as a dynamical system - only the relative motion of energy/matter has any ontological meaning: this has been the main theme of spacetime modeling from the beginning.  In dealing with this failure of the FLRW-GR dynamics we now show that a judicious choice of  $\Omega_\Lambda$ and  $\Omega_m$ can mock up the 3-space expansion, but only by introducing an extraneous and spurious acceleration.

\section{Predicting the $\Lambda$CDM  Parameters  $\Omega_\Lambda$ and $\Omega_{DM}$}

\begin{figure}
\vspace{0mm}
\hspace{-2mm}\includegraphics[scale=0.5]{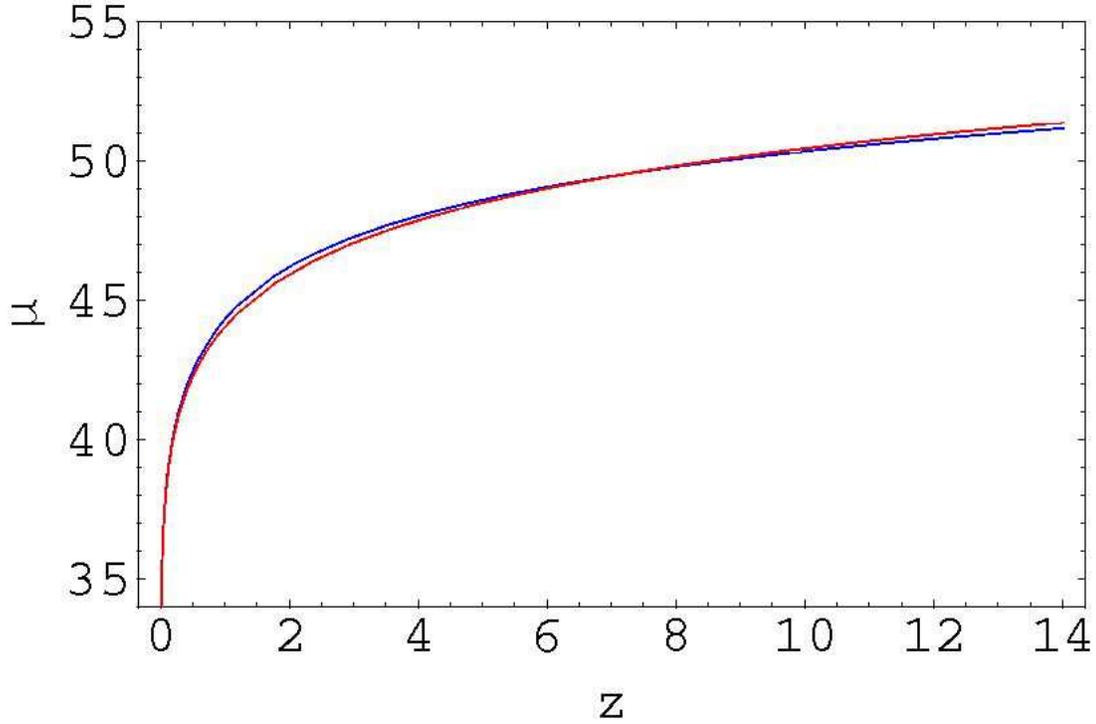}
\vspace{-1mm}\caption{\small{ Comparison of $\Lambda$CDM distance modulus $\mu(z)$ with $\Omega_\Lambda=0.73, \Omega_m=1-\Omega_\Lambda=0.27$, (blue plot), with the distance modulus from  the uniformly-expanding dynamical 3-space (red plot). The small difference, which could not be distinguished by the observational data, over this redshift range, demonstrates that the FLRW-GR model simulates the parameter-free uniformly-expanding dynamical 3-space prediction.  Hence  the `standard model'  values $\Omega_\Lambda=0.73$, $\Omega_m=0.27$ are predictable without reference to the actual supernovae/GRB magnitude-redshift data - there is no need to invoke `dark energy' nor `dark matter'. The FLRW-GR model does not permit an expanding space unless there is energy residing in the space.}
\label{fig:difference}}\end{figure}

It is argued herein that `dark energy' and `dark matter' arise in the FLRW-GR cosmology because in that model space cannot expand unless there is an energy density present in the space, if that space is flat and the energy density is pressure-less. Then essentially fitting the Friedmann  model  $\mu(z)$ to the dynamical 3-space cosmology $\mu(z)$ we obtain 
 $\Omega_\Lambda=0.73$, and so $\Omega_m=1-\Omega_\Lambda=0.27$. These values arise from a best fit for $z\in\{0,14\}$, and the quality of the fit is shown in figure \ref{fig:difference}.  The actual values for $\Omega_\Lambda$ depend on the red-shift range used, as the Hubble functions for the FLRW-GR and dynamical 3-space have different functional dependence on $z$. These values are of course independent of the actual observed redshift data. In fitting the Friedmann dynamics to the supernovae/GRB magnitude-redshift data the  best fit is  $\Omega_\Lambda=0.73$, and so $\Omega_m=0.27$ \cite{DETask}, p40. Of course since this amount of matter is much larger than the observed baryonic matter, it is claimed that most of this matter is  the so-called `dark matter'. Essentially the current standard model of cosmology $\Lambda$CDM  is excluded from modelling a uniformly expanding dynamical 3-space, but by choice of the parameter $\Omega_\Lambda$ the Hubble function $H_F(z)$ can be made to fit the data. However $H_F(z)$ has the wrong functional form; when applied to the future expansion of the universe the Friedmann dynamics  produces a spurious  exponentially expanding universe.

\section{Implications of the Supernovae and Gamma-Ray-Burst Data}

As already noted above the supernovae and gamma-ray-burst data show that the universe is uniformly expanding, and that such an expansion  cannot be produced by the Friedmann GR dynamics  for a flat 3-space except by a judicious choice of the parameters $\Omega_\Lambda$ and $\Omega_m=1-\Omega_\Lambda$.   Nevertheless we find that the FLRW flat 3-space spacetime metric is relevant but that it does not satisfy the Friedmann equations. We shall now illustrate this by comparing the distance moduli from various choices of the density parameters in (\ref{eqn:H2}).  We consider four choices of parameter values  with  the plots shown in  Figs. \ref{fig:SN1} and \ref{fig:SN2}:

 (i) A pure `dark energy' or cosmological constant driven expansion has $\Omega_m=0,  \Omega_r=0, \Omega_\Lambda=1, \Omega_s=0$.  This produces  a Hubble plot that causes too rapid an expansion, and indeed an exponential expansion at all epochs. This choice fails to fit the data.
 
 (ii)  A matter only expansion  has
$\Omega_m=1,  \Omega_r=0, \Omega_\Lambda=0, \Omega_s=0$. This produces a Hubble expansion that is de-accelerating and fails to fit the data.

 (iii)  The $\Lambda$CDM  Friedmann-GR parameters are  $\Omega_m=0.27,  \Omega_r=0, \Omega_\Lambda=0.73 , \Omega_s=0$. They arise from a fit to the dynamical 3-space uniformly-expanding  prediction as well as a best fit to the observational data. This shows that the data is implying a uniformly expanding 3-space. The Friedmann equations demand that $\Omega_s=0$ in the pressure-less dust case.
   
  (iv) The zero-energy dynamical 3-space has $\Omega_m=0,   \Omega_r=0,  \Omega_\Lambda=0,  \Omega_s=1$, as noted above. The spatial expansion dynamics alone gives a good account of the data. The data cannot distinguish between cases (iii) and (iv). 
  
  Of course the EM radiation term $\Omega_r$ is non-zero but small and determines the expansion during the baryongenesis initial phase, as does the spatial dynamics expansion term because of the $\alpha$ dependence.
  
  \section{Age of Universe and WMAP  Data}

The age of the universe is of course theory dependent.  From (\ref{eqn:R4}) it is given in general  by 
\begin{equation}
t_0=\int_0^1\frac{da}{\dot{a}(t)}=\int_0^\infty
\frac{dz}{(1+z)H(z)}
\label{eqn:Age}\end{equation} 
and so we must choose a form for $H(z)$, and one that models the redshift back to the Big Bang ($z=\infty$).  However we only have, at best, knowledge of $H(z)$ back to say $z \approx 7$. The FLRW-GR $H(z)$ essentially  fits to the 3-space form for $H(z)$ over a considerable range of $z$ values, as shown in figure \ref{fig:difference},  but not over the full $z$-range as shown in 
figure \ref{fig:Rtplot}. Indeed figure  \ref{fig:Rtplot} shows that the two $a(t)$ functions do differ, but that nevertheless they give essentially the same age for the universe. This is just an accident.  However as noted when applied to the future expansion  another extrapolation is employed and the FLRW-GR model predicts an exponential expansion, while the 3-space dynamics model predicts a continuing uniform expansion. From (\ref{eqn:H2a}), with $\alpha=0$, we obtain $t_0=1/H_0$.   However  there will be changes to this from including effects of baryonic matter and that when the universe is inhomogeneous $\rho_{DM}$ may not be small or even positive, and would not evolve as conserved matter does as in (\ref{eqn:evolve}).

 Analysis of the CMB anisotropies by WMAP \cite{WMAP1, WMAP2,Kosowsky}  have given results that are consistent with the $\Lambda$CDM model. However as noted herein that model involves a Hubble function that can also be matched by the Hubble function from the dynamical 3-space.  So the {\it concordance} between  fitting  the supernovae/GRB data and the CMB data to the $\Lambda$CDM model does not imply the correctness of   this model.  This issue has been discussed by Efstathiou and Brown \cite{Efstathiou}, and is known as the geometric degeneracy effect.  What is most telling in this context is more than the existence of this degeneracy effect, but that the $\Lambda$CDM model
 parameters can be accurately computed without reference to the observational data, so they are purely artifacts of using the FLRW-GR $\Lambda$CDM model.
 
   In this context we also note another geometric degeneracy, namely that if we use a FRW metric with a non-flat 3-space then the Friedmann equations now permit the term  with coefficient $b$ in  (\ref{eqn:R3}), but with $\alpha=0$, arises.   This term, however, has completely different origins: in the FLRW-GR cosmology it is associated with 3-space curvature, while above it is related to the dynamics of the {\it flat} 3-space.  
 
 So from the beginning of cosmology the flawed Friedmann model of an expanding universe with a non-dynamical 3-space has been employed.   The neglect of the 3-space dynamics up to now means that other methods for studying the so-called `dark energy' and `dark matter' need to be re-investigated: these include Baryonic Acoustic Oscillations (BAO),  Galaxy Cluster Counting (GCC) and Weak Gravitational Lensing  (WGL) \cite{DETask}.   In particular BAO analysis will be affected by the $\alpha$-dynamics term in (\ref{eqn:E1}) which can produce significant effects when the system is inhomogeneous. Similarly the GCC and WGL are also  affected by this $\alpha$-dynamics.  These effects impact on the determination of the baryonic matter content and on the computed age of the universe.
 
 \section{Ricci Curvature from the  Dynamical 3-Space}
 
 We now note the form of the Ricci scalar, which is a measure of the non-flatness of the induced spacetime metric. From either 
(\ref {eqn:PGmetric})  or (\ref{eqn:Hubblemetric}) we obtain the Ricci scalar to be
\begin{equation}
R=-6\left( \frac{{\dot{a}}^2}{a^2} +\frac{\ddot{a}}{a}\right) = \frac{-96+24\alpha}{(4+\alpha)^2t^2}\neq0
\label{eqn:RicciValue}\end{equation}
on using, say,  expression (\ref{eqn:spacexp}) for $a(t)$. So even though  the dynamical 3-space  leads to the FLRW  spacetime metric, with a flat 3-space, the spacetime itself is not flat.  Nevertheless it is important to note that the induced spacetime has no ontological significance - it is merely a mathematical construct.

\section{Conclusions}

The notion of dark energy and dark matter arose because in the analysis of the supernovae red-shift data \cite{S1,S2} Newtonian gravity was used in modelling the cosmological expansion of the universe, although usually presented in the more abstract formalism of the  FLRW-GR theory. Newtonian gravity
is only valid in special cases - such as outside of large spherical mass systems, such as the sun.  However a more general account of gravity requires an explicit  account of the dynamical 3-space, and the universality of this account has been established by using data from  bore-hole experiments, blackhole mass systematics in star systems ranging from globular clusters to large galaxies, light bending, spiral galaxy flat rotation curves, to the universe Hubble expansion.
The minimal model of  a classical dynamical 3-space requires  two-parameters, with one being $G$ and the other being $\alpha$. That this $\alpha$ is the fine structure constant is determined from various experimental/observational data.  Generalising the Schr\"{o}dinger and Dirac equations then explains the phenomenon of gravity -  gravity is an emergent phenomenon arising from the wave-nature of quantum matter.   The dynamical 3-space theory is then shown to explain various phenomena, including the so-called `dark matter' effects - essentially these are related to the  $\alpha$-dynamics that is missing from Newtonian gravity and GR.  The 3-space dynamics has an expanding flat-universe solution that gives a parameter-free account of the supernovae/GRB data.
This expansion occurs even when the energy density of the universe is zero. In contrast the FLRW-GR expansion dynamics only permits an expanding universe when the energy density, in the case of a pressure-less dust, is non-zero, and also essentially large.  To fit the expanding 3-space solution   a least-squares best-fit gives  $\Omega_\Lambda=0.73$ and $\Omega_m=0.27$ in the FLRW-GR model, independent of the observational data.  Not surprisingly these are the exact values found from fitting the FLRW-GR dynamics to the supernovae/GRB data.  However a spurious aspect to this is that the FLRW-GR fit  generates an anomalous exponential expansion in the future, as the FLRW-GR Hubble function has the wrong functional form.
Because of the dominance of  $\Omega_\Lambda=0.73$ and $\Omega_m=0.27$ the FLRW-GR dynamics has become known as the $\Lambda$CDM `standard' model of cosmology.   It is thus argued that the Friedmann dynamics for the universe has been flawed from the very beginning of cosmology, and that the new high-precision supernova data has finally made that evident.   The derived theory of gravity does away with the need for  `dark energy' and `dark matter'.   The Friedmann dynamics and its use as  the  $\Lambda$CDM standard model of cosmology has had a long  and tortuous evolution, but essentially it is Newton's theory of gravity applied to the whole universe, and so well beyond its established  regime.


\begin{thebibliography}{99}
\bibitem{Friedmann}  Friedmann, A. :   \"{U}ber die KrŸmmung des Raumes, Z. Phys.  { 10}, 377-386. (English translation in: 1999,  Gen. Rel. Grav.  31, 1991-2000.) (1922)
\bibitem{Lemaitre} Lema\^{i}tre. G.:   Expansion of the Universe, A Homogeneous Universe of Constant Mass and Increasing Radius Accounting for the Radial Velocity of Extra-Galactic Nebulae, Monthly Notices of the Royal Astronomical Society, { 91}, 483-490, (1931). Translated from Lema\^{i}tre G Un Univers Homog\`{e}ne de Masse Constante et de Rayon Croissant Rendant Compte de la Vitesse Radiale des N\'{e}buleuses Extra-Galactiques, { Annales de la Soci\'{e}t\'{e} Scientifique de Bruxelles}  { A47}, 49Ð56 (1927)
\bibitem{Robertson}  Robertson, H.P.:  { Kinematics and World Structure},  { Astron. J.}  { 82},  248-301;
1936,  {\b 83}, 187-200;  1936, {\bf 83}, 257-271 (1935)
\bibitem{Walker} Walker A.G.:  { On Milne's Theory of World-Structure}, {Proc.  London Math. Soc.}  {\bf 2 42}, 90-127 (1937)
\bibitem{PS}  Perlmutter, S. and Schmidt, B.P.:   { Measuring Cosmology with Supernovae}, in { Supernovae and Gamma Ray Bursters}, (Weiler K, Ed., Springer, Lecture Notes in Physics), { 598}, 195-217 (2003)
\bibitem{Book} Cahil,l  R.T.:  { Process Physics: From Information Theory to Quantum Space
       and Matter},  (Nova Science Pub., New York) (2005)
 \bibitem{S1} Riess, A.G.: {\it et al.}:  { Astron. J.} { 116}, 1009 (1998)
\bibitem{S2} Perlmutter. S., {\it et al.}:   {\it Astrophys. J.} {\bf 517}, 565 (1999)
\bibitem{Dynamicalspace}  Cahill, R.T.:  {\it Dynamical 3-Space: A Review}, arXiv:0705.4146.
\bibitem{Cahillflyby} Cahill, R.T.:  {\it  Resolving Spacecraft Earth-Flyby Anomalies with Measured Light Speed Anisotropy}, {Progress in Physics} 2, 103-110 (2008)
\bibitem{Schrod} Cahill, R.T.:   {  Dynamical  Fractal  3-Space and the Generalised Schr\"{o}dinger  
Equation: Equivalence Principle and  Vorticity Effects},   { Progress in Physics},  {1}, 27-34 (2006)
\bibitem{Newton}  Newton, I.:  { Philosophiae Naturalis Principia Mathematica} (1687)
\bibitem{alpha}  Cahil,l R.T.:    { Gravity, `Dark Matter' and the Fine Structure Constant}, { Apeiron}, {
12}(2), 144-177 (2005)
 \bibitem{DM}   Cahill, R.T.:    {  `Dark Matter' as a Quantum Foam In-flow Effect}, in {
Trends in Dark Matter Research},  96-140,  (ed.  Val Blain J, Nova Science Pub., New York)     (2005 )
\bibitem{Ander89}  Ander, M.E.: {\it et al.}   { Test of Newton's Inverse-Square Law in the Greenland Ice Cap}, { Phys. Rev. Lett.},  { 62},  985-988 (1989)
\bibitem{Thomas90} Thomas, J. and Vogel, P.: { Testing the Inverse-Square Law of Gravity in Bore Holes at the Nevada Test Site}, { Phys. Rev. Lett.},   { 65},  1173-1176 (1990)
\bibitem{galaxies}  Cahill,   R.T.:  { Black Holes in Elliptical and Spiral Galaxies and in 
Globular Clusters}, { Progress in Physics}, { 3}, 51-56 (2005)
\bibitem{newBH} Cahill,  R.T.:   { Black Holes and Quantum Theory: The Fine Structure Constant Connection},  { Progress in Physics}, { 4}, 44-50 ( 2006)
\bibitem{URC} Persic, M., Salucci, P. and Stel, F,:   { The Universal Rotation Curve of Spiral Galaxies: I The Dark matter Connection}, Mon. Not. R. Astronom. Soc. { 281}, 27 (1996)
\bibitem{Hertz}  Hertz, H.:    { On the Fundamental Equations of Electro-Magnetics for Bodies in Motion}, { Wiedemann's Ann.}  { 41}, 369;  1962 { Electric Waves, Collection of Scientific Papers,}   (Dover Pub., New  York) (1890)
\bibitem{DMGalaxies}  Clowe, D. {\it et al.}:    {  Direct Experimental Proof of the Existence of Dark Matter}, {  Ap J},  {\bf 648}, L109 (C06) (2006)
\bibitem{JeeDMRing} Jee, M. {\it et al.}:  {Discovery of a Ring-Like  Dark Matter Structure in the Core of the Galaxy Cluster CL 0024+17}, { Ap. J.}, { 661}, 728-749 (2007)
\bibitem{DMRing}  Cahill,   R.T.:   { Dynamical 3-Space: Alternative Explanation of the `Dark Matter Ring'}, { 4}, 13-17 (2007)
\bibitem{Panleve} Panlev\'{e}, P.:   {  C.R. Acad. Sci.}, { 173}, 677 (1921)
\bibitem{Gullstrand} Gullstrand, A.:   {  Ark. Mat. Astron. Fys.},  {16}, 1 (1922)
\bibitem{GPB} Cahill,  R.T.:   { Novel Gravity Probe B Frame-Dragging Effect},  { Progress in Physics}, { 3}, 30-33 (2005)
\bibitem{data set} http://dark.dark-cosmology.dk/ $\sim$tamarad/SN/
\bibitem{Davis} Davis,  T., Mortsell, E., Sollerman, J. and  ESSENCE:  {  Scrutinizing Exotic Cosmological Models Using ESSENCE Supernovae Data Combined with Other Cosmological Probes},  astro-ph/0701510 (2007)
\bibitem{WV} Wood-Vassey,  W.M. {\it et al.}:   { Observational Constraints on the Nature of the Dark Energy: First Cosmological Results from the ESSENCE Supernovae Survey},  astro-ph/0701041 (2007)
\bibitem{Riess} Riess, A.G.  {\it et al.}:  {  New Hubble Space Telescope Discoveries of Type Ia Supernovae at  $z > 1$: Narrowing Constraints on the Early Behavior of Dark Energy}, astro-ph/0611572 (2007)
\bibitem{GRB} Schaefer, B.E.:  {  The Hubble Diagram to Redshift $ >$  6 from 69 Gamma-Ray Bursts},
{ Ap. J.} { 660}, 16-46 (2007)
\bibitem{WMAP1}  Bennet, C.L. {\it et al.}:   { Wilkinson Microwave Anisotropy Probe  (WMAP) Observations: Preliminary Maps and Basic Results}, { Astrophys. J. Suppl.}, { 148}, 1 (2003)
\bibitem{WMAP2}  Verde, L.  {\it et al.}:  {  First Year Wlikinson Microwave Anisotropy Probe (WMAP) Observations: Parameter Estimation Methodology},  {Astrophys. J. Suppl.}, { 148}, 195 ( 2003)
\bibitem{Kosowsky} Kosowsky, A. {\it et al.}:   {  Efficient Cosmological Parameter Estimation from Microwave Background Anisotropies}, { Phys. Rev. D}, { 66}, 063007 (2002)
\bibitem{Efstathiou} Efstathiou, G. and Bond, J.R.: {Cosmic Confusion: Degeneracies Among Cosmological Parameters Derived from Measurements of Microwave Background Anisotropies}, {Mon. Not. Roy. Astron. Soc.},  { 304}, 75-97 (1999)
\bibitem{DETask}  Albrecht, A. {\it et al.}:   {  Report of the Dark Energy Task Force},  arXiv:astro-ph/0609591v1 (2006)


\end{thebibliography}
\end{document}